\documentclass[aps,prb,reprint,twocolumn,superscriptaddress,floatfix,citeautoscript,longbibliography]{revtex4-2}
\usepackage[bookmarks=true,colorlinks,linkcolor=OrangeRed,urlcolor=NavyBlue,citecolor=RoyalBlue]{hyperref}
\usepackage{epsfig,amsmath,amssymb,color,comment,physics}
\usepackage[makeroom]{cancel}
\usepackage[caption=false,font=normalsize]{subfig}
\usepackage{mathrsfs}
\usepackage[countmax]{subfloat}
\usepackage[normalem]{ulem}
\usepackage[english]{babel}
\usepackage{dsfont}
\usepackage{placeins}
\usepackage[dvipsnames]{xcolor}
\usepackage{svg} 
\usepackage{graphicx}
\usepackage{appendix}
\usepackage{multirow} 
\usepackage{graphicx}  
\usepackage{bbold} 
\usepackage{dcolumn}
\usepackage{bm}
\usepackage{comment}
\usepackage[capitalise]{cleveref}

\graphicspath{ {./figs/} }

\begin{document}

\title{High-temperature series expansion of the dynamic Matsubara spin correlator}

\author{Ruben Burkard}
\affiliation{Institut für Theoretische Physik, Universit\"at T\"ubingen, Auf der Morgenstelle 14, 72076 T\"ubingen, Germany}

\author{Benedikt Schneider}
\affiliation{Department of Physics and Arnold Sommerfeld Center for Theoretical
Physics, Ludwig-Maximilians-Universit\"at M\"unchen, Theresienstr.~37,
80333 Munich, Germany}
\affiliation{Munich Center for Quantum Science and Technology (MCQST), 80799 Munich, Germany}

\author{Bj\"orn Sbierski}
\affiliation{Institut für Theoretische Physik, Universit\"at T\"ubingen, Auf der Morgenstelle 14, 72076 T\"ubingen, Germany}

\date{\today}

\begin{abstract}
The high-temperature series expansion for quantum spin models is a well-established tool to compute thermodynamic quantities and equal-time spin correlations, in particular for frustrated interactions. We extend the scope of this expansion to the dynamic Matsubara spin-spin correlator and develop an algorithm that yields exact expansion coefficients in the form of rational numbers. We focus on Heisenberg models with a single coupling constant $J$ and spin lengths $S\in\{1/2,1\}$. The expansion coefficients up to 12th order in $J/T$ are precomputed on all possible $\sim 10^6$ graphs embeddable in arbitrary lattices and are provided in a repository. This enables calculation of static momentum-resolved susceptibilities for arbitrary site-pairs or wavevectors. We test our results for the antiferromagnetic $S=1/2$ chain and triangular lattice model. An important application that we discuss in a companion letter is the calculation of real-frequency dynamic structure factors. This is achieved by identifying the high-frequency expansion coefficients of the Matsubara correlator with frequency moments of the spectral function.
\end{abstract}

\maketitle

\section{Introduction} \label{sec:intro}
The high-temperature series expansion (HTE) is an invaluable tool for the theoretical analysis of quantum spin systems in thermal equilibrium, for monographs see, e.g., Refs.~\onlinecite{dombPhaseTransitions1974,oitmaaSeriesExpansion2006}. The HTE is oblivious to frustration and entanglement, is formulated directly in the thermodynamic limit and remains applicable in high dimensions. Up to now, the HTE can target thermodynamic quantities like entropy, heat capacity or uniform susceptibility and was also applied to obtain equal-time spin correlation functions $G_{ii^\prime}^{zz}=\left\langle S_i^z S_{i^\prime}^z \right\rangle$ for arbitrary site-pairs $ii^\prime$. Key technical advancements of the HTE method included an extension to arbitrary spin-length $S$ \cite{schmidt_heisenberg_2001, lohmannTenthorderHightemperature2014}, flexible open-source software packages \cite{hehnHightemperatureSeries2017} or inclusion of magnetic fields \cite{pierre_high_2024}.
As the name suggests, the main challenge of HTE is to reach temperatures much below the spin interaction $T\ll J$ when resummation schemes of the bare series give ambiguous and thus non-reliable results. To some extent, these issues can be bypassed if qualitative information on the very low-$T$ behavior  is available \cite{bernu_specific_2001}.

HTE has an impressive track-record of achievements: For example, it has first revealed \cite{elstner_finite_1993} the \emph{anomalous} intermediate-$T$ behavior of the nearest-neighbor $S=1/2$ Heisenberg AFM on the triangular lattice which cannot be understood in the renormalized-classical picture suggested by the ordered ground state. The latter picture predicts a much smaller entropy and a much larger correlation length than found from HTE. The underlying physical reason of this effect is still not well understood \cite{ghioldi_dynamical_2018,chen_two-temperature_2019}. Another example is the accurate quantitative analysis of thermal phase transitions in otherwise challenging three-dimensional frustrated models \cite{gonzalez_finite-temperature_2023}.

However, in order to achieve a better understanding of collective phenomena arising in equilibrium quantum spin systems it is mandatory to also computationally target spin dynamics. Here, the most elementary observable is the dynamical spin-spin correlator $G^{zz}_{ii^\prime}(t)=\left\langle S_{i}^{z}(t)S_{i^{\prime}}^{z} \right\rangle$
with $S_{i}^{z}(t)=e^{iHt} S_i^z e^{-iHt}$ representing a local spin operator in the Heisenberg picture \cite{bruusManyBodyQuantum2004}. The dynamical structure factor which is defined as a spatial and temporal Fourier transform of $G^{zz}_{ii^\prime}(t)$ contains rich information on the dipolar excitation spectrum, the presence and stability of quasiparticles or the fractionalization of spin-flip excitations \cite{mourigal_fractional_2013}. It is routinely measured using inelastic neutron scattering \cite{khatua_experimental_2023} in the solid-state context and recently also for cold-atom quantum simulators \cite{prichard_observation_2025} via Raman spectroscopy. Although the dynamical structure factor can be calculated in various ways for different settings \cite{lacroixIntroductionFrustrated2011} (e.g.~via exact diagonalization and linked-cluster methods \cite{sherman_structure_2018}, spin-wave theory \cite{auerbachInteractingElectrons1994} and tensor-networks \cite{drescher_dynamical_2023}), it is still desirable to approach spin dynamics via the HTE and benefit from its above mentioned strengths. This is particular important for (strongly frustrated) spin liquid candidates \cite{savaryQuantumSpin2016} where the observables available from conventional HTE are often featureless as functions of temperature and momentum. A natural intermediate goal to this endeavor is the extension of the HTE for the dynamical spin correlation function in imaginary time $t \rightarrow -i \tau \in i \mathbb{R}$ which is a somewhat simpler task given the benefits of the Matsubara formalism for perturbation theory in equilibrium \cite{bruusManyBodyQuantum2004}. 

In this work, we develop this extension of the HTE to the dynamic imaginary frequency (Matsubara) spin-spin correlator. We term this method dynamic HTE (Dyn-HTE). We lay out the general formalism and apply it to Heisenberg models with a single coupling constant $J$ and spin lengths $S\in\{1/2,1\}$. The coefficients for the expansion in powers of $x\equiv J/T$ up to order $n_{max}=12$ (and in inverse frequency to order $\left\lfloor n_{max}/2\right\rfloor$) are precomputed exactly in the form of rational numbers on all possible $\sim 10^6$ graphs and are offered for download along with powerful tools for the creation of arbitrary lattices and efficient graph embeddings \cite{Dyn-HTEsoftware_v1_0}. 
On the technical side Dyn-HTE hinges on the exploitation of the recently developed Kernel trick \cite{halbingerSpectralRepresentation2023} which solves the $(n+2)$-fold imaginary-time integrals (required at expansion order $n$) analytically. Interestingly, the numerical cost for the expansion in order $n$ is only modestly increased compared to the conventional HTE for the equal-time correlator. Depending on graph topology the extra effort is by a factor of $n$ in the worst case. 

For historic context, early developments of spin-diagrammatic schemes date back to the late 1960s \cite{vaksSpinWaves1968,izyumovStatisticalMechanics1988} where the Wick theorem was generalized for spin operators, see also Ref.~\cite{gollSpinFunctional2019} for a modern recursive formulation. Recently, these ideas have partially been revived in the development of a functional-renormalization group approach for spin systems \cite{kriegExactRenormalization2019,tarasevychDissipativeSpin2021,gollSpinFunctional2019} which in principle can also be used to generate order-by-order expansions of the Matsubara correlator \cite{ruckriegel_recursive_2024}. However, the high expansion orders achieved by Dyn-HTE in this work have not been matched by any other approach.

One experimental observable directly available from Dyn-HTE is the static susceptibility for arbitrary site-pairs. We consider the AFM $S=1/2$ Heisenberg chain and compare the static susceptibility from Dyn-HTE against error controlled quantum Monte Carlo (QMC). We also report the static susceptibility for the frustrated triangular lattice model. As an application we employ the static susceptibility of Dyn-HTE to showcase the accuracy of a simple approximate parametrization of its momentum dependence as suggested recently under the name renormalized mean-field form.  

Finally, as motivated above, a main application of the Matsubara correlator is its analytical continuation to the real-frequency dynamical spin structure factor \cite{bruusManyBodyQuantum2004}. A diverse set of methods like QMC or pseudo-fermion based diagrammatic approaches \cite{sandvikComputationalStudies2010,kulagin_bold_2013,reuther_j1_2010,niggemann_frustrated_2021,muller_pseudo-fermion_2024} produce approximate numerical correlator data on a limited set of points on the Matsubara axis. In such a situation, despite recent advances \cite{fei_nevanlinna_2021,nogaki_bosonic_2023,huang_barycentric_2025,kliczkowski_autoencoder-based_2024,shao_progress_2023}, this analytical continuation is an ill-conditioned and error-prone procedure. For Dyn-HTE, in contrast, the obtained exact expansion can be regrouped in form of a high-frequency expansion (in inverse Matsubara frequency). In our companion work \cite{burkard_DynHTE_letter} we show that the associated expansion coefficients can be identified with the frequency moments of the (real-frequency) spectral function from which the dynamical structure factor can be reconstructed by standard methods. This means that Dyn-HTE allows to bypass the standard ill-defined analytical continuation procedure.

\section{Heisenberg spin model} \label{sec:model}

We consider a system of length-$S$ quantum spins with operators $S^\alpha_i$ where $\alpha = x,y,z$. The subscript $i=1,2,...,N$ refers to the site at position $\mathbf{r}_i$ of an arbitrary lattice $\mathcal{L}$. The spins interact via Heisenberg exchange characterized by a \emph{single} coupling constant $J$ along an arbitrary subset of all $N(N-1)/2$ site pairs which we call bonds $(ii^\prime)$. This includes the important case of symmetry-related nearest-neighbor interactions, but also all-to-all or spatially disordered (but equal) interactions like in a lattice with vacancies. The Hamiltonian reads
\begin{equation}
    H=J\sum_{(ii^\prime)}\left(S_{i}^{+}S_{i^\prime}^{-}+S_{i}^{-}S_{i^\prime}^{+}+S_{i}^{z}S_{i^\prime}^{z}\right)
    \equiv
    J\sum_{(ii^\prime)} V_{ii^\prime}, \label{eq:H}
\end{equation}
where $i<i^\prime$ (no on-site terms). The spin ladder operators are $S_i^{\pm}=\left( S_i^{x}\pm iS_i^{y} \right) /\sqrt{2}$.  We further assume the absence of external magnetic fields or spontaneously broken symmetries (time-reversal and spin-rotation). None of these assumptions or the restriction to the model in Eq.~\eqref{eq:H} are fundamental for Dyn-HTE and can be relaxed in future extensions in parallel to the developments in the history of conventional HTE \cite{oitmaaSeriesExpansion2006}. 

We assume thermal equilibrium at temperature $T=1/\beta$ ($k_B=\hbar=1$). The density matrix is $\rho = e^{-\beta H} / Z$ with $Z=\mathrm{tr} \, e^{-\beta H}$ the partition function and thermal averages of operators are given by
\begin{equation}
 \left\langle ...\right\rangle \equiv \mathrm{tr}[...\rho]=\mathrm{tr}[...e^{-\beta H}]/Z. \label{eq:def<...>}   
\end{equation}
For most applications, we assume translational invariance and place the spins on a regular lattice at positions
\begin{equation}
    \mathbf{r}_i=\mathbf{R}_i+\mathbf{b}_{i} \label{eq:r=R+b}
\end{equation}
where the label $i$ selects both the site $\mathbf{R}_i$ of a Bravais lattice and a basis vector $\mathbf{b}_i \in \{\mathbf{b}^{(1)},...,\mathbf{b}^{(N_b)}\}$. We assume $N_c$ Bravais lattice sites (with periodic boundary conditions) and $N_b$ basis vectors so that $N=N_c N_b$. 

\section{Matsubara spin correlator and Dyn-HTE} \label{sec:MatsubaraG}

The imaginary-frequency (Matsubara) spin-spin correlator is defined as \cite{bruusManyBodyQuantum2004}
\begin{equation}
G^{zz}_{ii^\prime}(i\nu_{m})=T\int_{0}^{\beta} \!\! \mathrm{d}\tau\mathrm{d}\tau^{\prime}\,e^{i\nu_{m}(\tau-\tau^{\prime})}\,
\left\langle \mathcal{T}S_{i}^{z}(\tau)S_{i^{\prime}}^{z}(\tau^{\prime})\right\rangle
,\label{eq:G(nu_m)}
\end{equation}
where $\nu_m=2 \pi T m$ with $m\in \mathbb{Z}$ is a (bosonic) Matsubara frequency and $S_{i}^{z}(\tau)= e^{H\tau}S_{i}^{z}e^{-H\tau}$ denotes \emph{imaginary} time evolution. The operator
$\mathcal{T}$ enforces imaginary time ordering, the operator with larger imaginary time argument is moved to the left. Note that the double integral is often replaced by a single integral over the difference $\tau-\tau^\prime$ but the form in \eqref{eq:G(nu_m)} will prove useful later. Due to spin rotation symmetry in the Heisenberg Hamiltonian, Eq.~\eqref{eq:G(nu_m)} also determines correlations for all other spin-flavor combinations, $G^{\alpha \alpha^\prime}=\delta_{\alpha \alpha^\prime}G^{zz}$ and the $zz$ superscript of $G^{zz}$ is dropped in the following to ease notation. The spatial Fourier transform of Eq.~\eqref{eq:G(nu_m)} reads
\begin{equation}
    G_{\mathbf{k}}(i\nu_{m})=\frac{1}{N}\sum_{i,i^{\prime}}e^{-i\mathbf{k}\cdot(\mathbf{r}_{i}-\mathbf{r}_{i^{\prime}})}G_{ii^{\prime}}(i\nu_{m}). \label{eq:G(mu_m)_k}
\end{equation} 

The symmetries of $G_{ii^\prime}(i\nu_{m})$ follow from its definition \eqref{eq:G(nu_m)} and the Hamiltonian \eqref{eq:H}:
Hermitian conjugation leads to a reality condition $G_{ii^\prime}(i\nu_{m}) \in \mathbb{R}$ and time-reversal symmetry ensures invariance under frequency flip $G_{ii^\prime}(i\nu_{m})=G_{ii^\prime}(-i\nu_{m})$. This also implies symmetry under exchange of site-indices $G_{ii^\prime}(i\nu_{m})=G_{i^{\prime}i}(i\nu_{m})$ which leads to 
\begin{equation}
G_\mathbf{k}(i\nu_m)=G_{-\mathbf{k}}(i\nu_m)^\star.   \label{eq:Gk=G-k*}
\end{equation}
Hence, with inversion symmetry, $G_\mathbf{k}(i\nu_m)\in \mathbb{R}$.

As we show in Sec.~\ref{sec:derivationDyn-HTE}, the Dyn-HTE of the Matsubara correlator \eqref{eq:G(nu_m)} takes the form of a double expansion in $x=J/T$ and inverse Matsubara frequency $1/ \nu_m$. To order $n_{max}$ in $x$, Dyn-HTE reads
\begin{eqnarray}
&&TG_{ii^{\prime}}(i\nu_{m})= \nonumber \\
& & \begin{cases}
p_{ii^{\prime}}^{(0)}(x) \!\!&\!:\!m=0\\
\sum_{r=1}^{r_{max}}p_{ii^{\prime}}^{(2r)}(x)\left(\frac{x}{2\pi m}\right)^{2r} \!\!&\!:\!m\neq0
\end{cases}\!+O(x^{n_{max}+1}),
\label{eq:Dyn-HTE}   
\end{eqnarray}
where $r_{max}=\left\lfloor n_{max}/2\right\rfloor$ is the integer floor of $n_{max}/2$. The dimensionless polynomials $p_{ii^{\prime}}^{(2r)}(x)$ for are of degree $n_{max}-2r$ with (real) rational coefficients, here $r=0,1,2,...$. Depending on the lattice and site-pair $ii^{\prime}$, some of the polynomial coefficients can be zero. Note that the form \eqref{eq:Dyn-HTE} is consistent with the symmetries of the Matsubara correlator discussed above.

Our open-source numerical implementation \cite{Dyn-HTEsoftware_v1_0} provides the exact polynomial coefficients in Eq.~\eqref{eq:Dyn-HTE} as rational numbers for \emph{arbitrary} lattices $\mathcal{L}$ and all site-pairs $ii^\prime$ therein. Currently spin lengths $S\in\{1/2,1\}$ and maximum expansion order $n_{max}=12$ are available.   

We conclude this section with a review of the physical content \cite{bruusManyBodyQuantum2004} of the Matsubara spin correlator \eqref{eq:G(nu_m)}. At zero frequency, $m=0$, the Matsubara correlator yields the static susceptibility $\chi_{ii^\prime}$. The latter is defined as the (negative) linear isothermal response of $z$-magnetization at site $i$ to a static local magnetic field perturbing the Hamiltonian $H$ via $h^z_{i^\prime} S_{i^\prime}^z$ \cite{kwok_correlation_1969},
\begin{equation}
   G_{ii^{\prime}}(i\nu_{m}=0) = \chi_{ii^\prime} = -\partial_{h^z_{i^{\prime}}}\left\langle S_{i}^{z}\right\rangle|_{h_{i^{\prime}}^z=0}. \label{eq:chi_isothermal}
\end{equation}

At finite frequency $m \!\neq\! 0$, the Matsubara correlator does \emph{not} directly represent an observable physical quantity. It is mainly considered for the simplicity of the diagrammatic field-theoretical framework which can be employed for its computation \cite{bruusManyBodyQuantum2004}. Usually, numerical data for the Matsubara correlator $G_{ii^\prime}(i\nu_m)$ evaluated at a finite set of frequencies $\nu_m$ is linked to its real-frequency version by analytic continuation as discussed in Sec.~\ref{sec:intro}. Dyn-HTE allows to take a different and more stable route and never evaluates numerical values for $G_{ii^\prime}(i\nu_m)$: As we show in our companion letter \cite{burkard_DynHTE_letter}, the high-frequency expansion coefficients from Eq.~\eqref{eq:Dyn-HTE} are in direct correspondence to the short-time expansion coefficients of the real-time correlator,  
\begin{equation}
 p_{ii^\prime}^{(2r)} \sim   \partial_{t}^{2r-1}
\!\!
\left\langle S_{i}^{z}(t)S_{i^{\prime}}^{z}\right\rangle \! |_{t=0}.  \label{eq:pToMoments}
\end{equation}
The object on the right-hand side of Eq.~\eqref{eq:pToMoments} can also be identified with the (real-frequency) moments of the spectral function \cite{viswanath_recursion_1994}. Using equations of motion, the moments can be unraveled as equal-time correlators of $2r-1$ fold nested commutators, $\! \sim \! \left\langle [...[S_{i}^{z},\!H],\!H],...,\!H]S_{i^{\prime}}^{z}\right\rangle$. Even for moderate $r$, these are hard to compute by standard means beyond the case of one dimension and $T=\infty$. This explains the value of Dyn-HTE's access to (the HTE of) $p_{ii^\prime}^{(2r)}$ for these cases. The full spectral function and dynamical structure factor can be reconstructed based on a moderate number of frequency moments via continued fraction representations \cite{mori_continued-fraction_1965,viswanath_recursion_1994,pairault_strong-coupling_2000,perepelitsky_transport_2016,bhattacharyyaMetallicTransportHardcore2024}. In our companion letter \cite{burkard_DynHTE_letter} we discuss this in detail and show benchmarks and applications centered around the dynamical structure factor. In contrast, the remainder of the current work will be concerned with the efficient evaluation of the Dyn-HTE, i.e.~the polynomials $p_{ii^\prime}^{(2r)}(x)$ in Eq.~\eqref{eq:Dyn-HTE}. Here we also present, benchmark and discuss results for the Matsubara correlator in imaginary frequency. 

\section{Derivation of Dyn-HTE} \label{sec:derivationDyn-HTE}

In this central technical section we derive the Dyn-HTE as in Eq.~\eqref{eq:Dyn-HTE} and provide an efficient algorithm for its calculation. To start, we split the Hamiltonian \eqref{eq:H} into a non-interacting part $H_0=0$ (vanishing in the absence of a magnetic field) and an interacting part $V$ (here the full $H$) and define the dimensionless expansion coefficients $c_{ii^\prime}^{(n)}(i\nu_{m})$ of the Matsubara correlator via
\begin{equation}
    TG_{ii^\prime}(i\nu_{m}) = \sum_{n=0}^\infty (-x)^n c_{ii^\prime}^{(n)}(i\nu_{m}),\;\;\;\;\;\;\;(x=\frac{J}{T}).
    \label{eq:TG-expansion}
\end{equation}
The lowest-order $n=0$ term represents the Curie susceptibility of a free spin which is local and static (non-zero only for $m=0$), $c_{ii^\prime}^{(0)}(i\nu_{m})=\delta_{ii^\prime}\delta_{0,m}S(S+1)/3$. 

\subsection{Expansion coefficients}

General perturbation theory for the Matsubara correlator \cite{bruusManyBodyQuantum2004} provides an expression for the remaining expansion coefficients in Eq.~\eqref{eq:TG-expansion} at order $n=1,2,...$
\begin{widetext}
\begin{equation}
c_{ii^\prime}^{(n)}(i\nu_{m}) = \frac{T^{n+2}}{n!}\sum_{b_{1},...,b_{n}}\int_{0}^{\beta}e^{i\nu_{m}(\tau-\tau^{\prime})}\mathrm{d}\tau_{1}...\mathrm{d}\tau_{n}\mathrm{d}\tau\mathrm{d}\tau^{\prime}\left\langle \mathcal{T}V_{b_{1}}(\tau_{1})...V_{b_{n}}(\tau_{n})S_{i}^{z}(\tau)S_{i^{\prime}}^{z}(\tau^{\prime})\right\rangle _{0}^{\mathrm{V-con}}.
\label{eq:c(m)^(n)}
\end{equation}
The sum is over bonds $b_l=(i_l i_l^\prime)$ and operator time arguments as in $V_{b_{1}}(\tau_{1})$ now refer to (interaction-picture) evolution with respect to $H_0$ only. This is trivial for our case, $H_0=0$, e.g.~$V_{b_{1}}(\tau_{1})=V_{b_1}$. Nevertheless we must keep the time arguments because by virtue of $\mathcal{T}$ they determine the ordering of the string of generally non-commuting operators. These operators are thus distinguishable regardless of the chosen bonds $b_l$. The $\left\langle ...\right\rangle _{0}$ denotes a thermal average with respect to Hamiltonian $H_0$. Since $H_0=0$, the former is simply an (effectively) infinite temperature average or normalized trace over the whole $N$-site spin Hilbert-space,
$\left\langle ...\right\rangle_0=\mathrm{tr}[...]/(2S+1)^{N} $
which factorizes according to site-index, e.g.~$\left\langle S_1^zS_2^+S_1^z S_2^-\right\rangle_0 = \left\langle S_1^zS_1^z\right\rangle_0 \left\langle S_2^+ S_2^-\right\rangle_0$ with the on-site operator order maintained. 

The superscript ``$\mathrm{V-con}$" on the right-hand side of Eq.~\eqref{eq:c(m)^(n)} refers to the $V$-connected correlator which contains all contributions to order $V^n$, also those from the expansion of the partition function $Z$ in the denominator on the right-hand side of Eq.~\eqref{eq:def<...>}. Following Ref.~\onlinecite{rossiDeterminantDiagrammatic2017}, it can be written via recursive subtractions from the ''full" correlator [without $\mathrm{V-con}$, c.f.~\eqref{eq:def<...>}],
\begin{eqnarray}
    \left\langle \mathcal{T}V_{b_{1}}(\tau_{1})...V_{b_{n}}(\tau_{n})S_{i}^{z}(\tau)S_{i^{\prime}}^{z}(\tau^{\prime})\right\rangle _{0}^{\mathrm{V-con}}
    &=&
    \left\langle \mathcal{T}V_{b_{1}}(\tau_{1})...V_{b_{n}}(\tau_{n})S_{i}^{z}(\tau)S_{i^{\prime}}^{z}(\tau^{\prime})\right\rangle _{0} \label{eq:Vcon} \\
    &-& \!\!
    \sum_{S\subsetneq\{1,...,n\}}
    \left\langle \mathcal{T}\left[\prod_{k\in S}V_{b_{k}}(\tau_{k})\right]S_{i}^{z}(\tau)S_{i^{\prime}}^{z}(\tau^{\prime})\right\rangle_{0}^{\mathrm{V-con}}
    \left\langle \mathcal{T}\prod_{l\in\{1,..,n\}\backslash S}V_{b_{l}}(\tau_{l})\right\rangle _{0}, \nonumber
\end{eqnarray}
where the sum is over true subsets $S$ of the set $\{1,...,n\}$ (including the empty set) and the recursion terminates at 
$
    \left\langle \mathcal{T}S_{i}^{z}(\tau)S_{i^{\prime}}^{z}(\tau^{\prime})\right\rangle_{0}^{\mathrm{V-con}} \equiv \left\langle \mathcal{T}S_{i}^{z}(\tau)S_{i^{\prime}}^{z}(\tau^{\prime})\right\rangle_{0}.
$
We insert Eq.~\eqref{eq:Vcon} in Eq.~\eqref{eq:c(m)^(n)} and obtain
\begin{eqnarray}
    c_{ii^{\prime}}^{(n)}(i\nu_{m})	&=&	\sum_{b_{1},...,b_{n}}\frac{T^{n+2}}{n!}\int_{0}^{\beta}e^{i\nu_{m}(\tau-\tau^{\prime})}\mathrm{d}\tau_{1}...\mathrm{d}\tau_{n}\mathrm{d}\tau\mathrm{d}\tau^{\prime} \label{eq:c(n)_ii'(m)-rec}\\
	&\times&
    \{ \left\langle \mathcal{T} V_{b_{1}}(\tau_{1})...V_{b_{n}}(\tau_{n})S_{i}^{z}(\tau)S_{i^{\prime}}^{z}(\tau^{\prime})\right\rangle _{0}
    - \!\!\!\!\!
    \sum_{S\subsetneq\{1,...,n\}} \!\! \left\langle \mathcal{T} \!\! \left[\prod_{k\in S}V_{b_{k}}(\tau_{k})\right] \! S_{i}^{z}(\tau)S_{i^{\prime}}^{z}(\tau^{\prime}) \!\! \right\rangle _{0}^{\!\!\mathrm{V-con}}
    \!\!  \left\langle \mathcal{T} \!\!\!\!\! \prod_{l\in\{1,..,n\}\backslash S} \!\!\!\!\! V_{b_{l}}(\tau_{l})\right\rangle _{\!\!0} \! \}. \nonumber
\end{eqnarray}
\end{widetext}

\subsection{Graph-based evaluation of bond-sums}

Compared to the established HTE for equal-time spin-spin correlators $\left\langle S_{i}^{z}S_{i^{\prime}}^{z}\right\rangle$, the new and challenging aspect in Eq.~\eqref{eq:c(n)_ii'(m)-rec} are the $n+2$-dimensional imaginary time integrals. We relegate the integral evaluation to section \ref{subsec:KernelTrick}. Here we first focus on the bond-sums in Eq.~\eqref{eq:c(n)_ii'(m)-rec}, which we expose by summarizing the rest as follows,
\begin{equation}
    c_{ii^{\prime}}^{(n)}(i\nu_{m}) \equiv	\sum_{b_{1},...,b_{n}}F^{(n)}_{ii^{\prime}}\left(b_{1},...,b_{n};i\nu_{m}\right). \label{eq:bond_sum_F}
\end{equation}
Efficient evaluation strategies for these bond-sums are well developed in the literature on conventional HTE \cite{oitmaaSeriesExpansion2006}. The following discussion of the necessary elements of these strategies is self-contained and does not assume any prior knowledge. In a nutshell, our plan is to evaluate the expansion of the Matsubara correlator not directly for a full lattice $\mathcal{L}$, but first for individual lattice snippets with only $n$ bonds which are called graphs, $g^{(n)}$. These are then added up as they fit into the full lattice. Formally speaking, we re-organize the sum over (potentially identical) lattice bonds $\sum_{b_1,...,b_n}$ as a sum over evaluations of $F^{(n)}_{ii^\prime}$ on these graphs,
\begin{eqnarray}
    c_{ii^\prime}^{(n)}(i\nu_m)  &=& \sum_{g^{(n)}} e(\mathcal{L},i,i^{\prime},g^{(n)})  c_{g^{(n)}}(i\nu_m), \label{eq:embedding}  \\
    c_{g^{(n)}}(i\nu_m) 
    &=& \!\!\!\!\!\!
    \sum_{\{e_{1},...,e_{n}\}\rightarrow\{\tilde{e}_{1},...,\tilde{e}_{n}\}}
    \!\!\!\!\!\!\!
    F^{(n)}_{jj^{\prime}}\left(\tilde{e}_{1},...,\tilde{e}_{n};i\nu_{m}\right).  \label{eq:c_g}
\end{eqnarray}

\begin{figure*}
\begin{centering}
\includegraphics[width=1\textwidth]{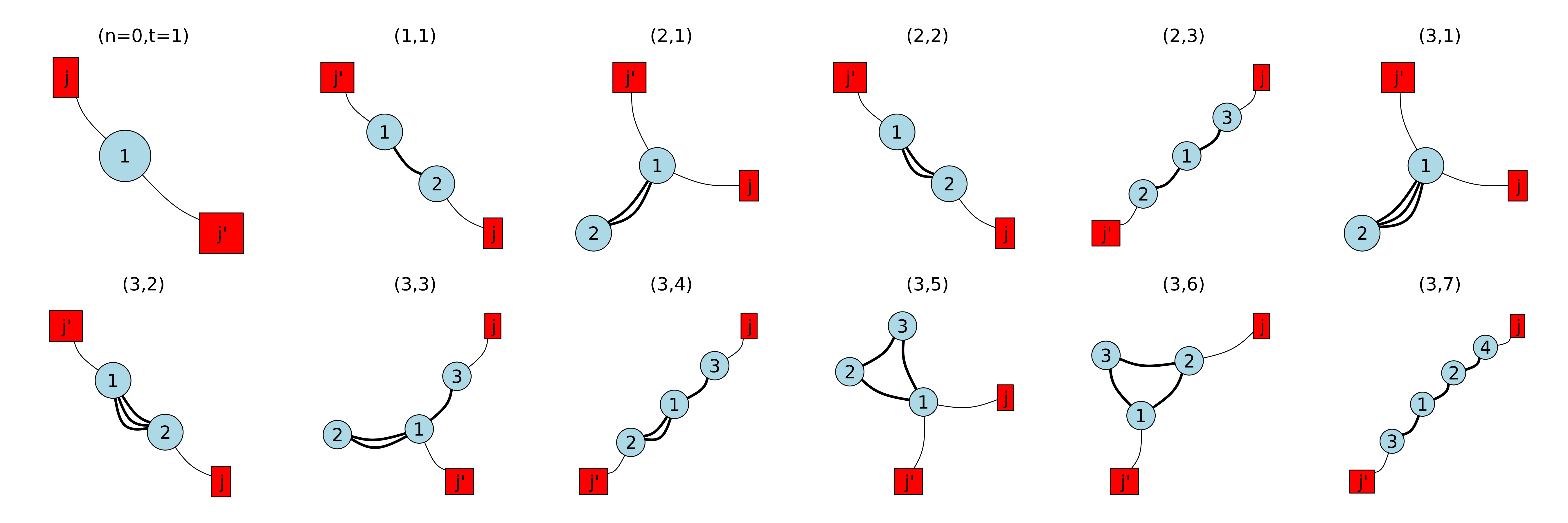}
\par\end{centering}
\caption{\label{fig:GraphsG}
All required graphs $g^{(n)}_t$ with $n=0,1,2,3$ edges (black lines) and arbitrary numbered vertices (blue circles). Terminal vertices $j,j^\prime$ (also blue circles) are indicated by their red-square terminal flags attached with thin gray lines. The symmetry factor is $s[g^{(n)}_t]=1$ for all graphs shown except for $(n,t)=(3,5)$ where it is two (exchange of vertices $2 \leftrightarrow 3$).
}
\end{figure*}

A (multi-)graph of order $n$ denoted by $g^{(n)}$ can be thought of as a lattice-snippet with $n$ bonds, see Fig.~\ref{fig:GraphsG}. Formally, it is defined by the multi-set of $n$ (not necessarily distinct) edges $\{e_{1},...,e_{n}\}$ (thick black lines) connecting its arbitrarily numbered vertices (blue circles) which contain the terminal vertices $j,j^\prime$ (blue circles highlighted by attached red-square flags). The terminal vertices indicate the position of the external operators $S_i^z$ and $S_{i^\prime}^z$.  Different graphs are distinguished by graph topology, Fig.~\ref{fig:GraphsG} shows all topological distinct graphs $g^{(n)}_t$, labeled by $t=1,2,...$ with up to $n=3$ edges that are required for our purpose. As we motivate later, two graphs only differing by the exchange of terminals $j \leftrightarrow j^\prime$ are considered topologically equivalent.

Having defined the graph $g^{(n)}$ as a lattice snippet, we next need to place it into the full lattice $\mathcal{L}$. This is done at the \emph{embedding} step, formally expressed in Eq.~\eqref{eq:embedding}. Here the embedding factor $e(\mathcal{L},i,i^{\prime},g^{(n)})\in\{0,1,2,...\}$ counts the number of sub-graph isomorphisms from the graph $g^{(n)}$ (with edge-multiplicities ignored) to the lattice $\mathcal{L}$ divided by the graphs' symmetry factor $s[g^{(n)}]\in \{1,2,...\}$ to avoid overcounting due to the arbitrary vertex numbering. The symmetry factor is the number of graph isomorphisms that keep the terminal vertices invariant and respect the edge-multiplicities, see the caption of Fig.~\ref{fig:GraphsG} for examples for $s[g^{(n)}]$.

For the embedding $g^{(n)} \rightarrow \mathcal{L}$ we require the assignments $(i,i^{\prime})\rightarrow(j,j^{\prime})$ or $(i,i^{\prime})\rightarrow(j^{\prime},j)$ to match the positions of the external operators. Taking into account also the latter assignment $(i,i^{\prime})\rightarrow(j^{\prime},j)$ allows us to skip graphs which are connected to another graph in the list by an exchange of $j\leftrightarrow j^{\prime}$. This assignment is not counted if $g^{(n)}$ is mapped to itself under $j\leftrightarrow j^{\prime}$. As an example, embedding factors for a number of simple lattices are given in Tab.~\ref{tab:e}. For two- and in particular three-dimensional lattices the embedding step can be numerically demanding. To ensure efficiency, our numerical implementation \cite{Dyn-HTEsoftware_v1_0} builds on lattice generating functionalities provided in the software package \emph{SpinMC.jl} \cite{buessen_fbuessenspinmcjl_2025}, relies on space-group symmetries for speed-up and uses advanced graph-theoretical algorithms \cite{noauthor_graphsjl_nodate, cordella_subgraph_2004}.
\begin{table}
    \centering
\begin{tabular}{c|c|c|c c c|c c c c c c c }
g & $\!g^{(0)}_1\!$ & $\!g^{(1)}_1\!$ & $\!g^{(2)}_1\!$ & $\!g^{(2)}_2\!$ & $\!g^{(2)}_3\!$ & $\!g^{(3)}_1\!$ & $\!g^{(3)}_2\!$ & $\!g^{(3)}_3\!$ & $\!g^{(3)}_4\!$ & $\!g^{(3)}_5\!$ & $\!g^{(3)}_6\!$ & $\!g^{(3)}_7\!$\tabularnewline
\hline 
dimer loc & 1 &  & 1 &  &  & 1 &  &  &  &   &  & \tabularnewline
dimer nn &  & 1 &  & 1 &  &  & 1 &  &  &   &  & \tabularnewline
\hline 
trimer loc & 1 &  & 2 &  &  & 2 &  &  &   & 1 &  & \tabularnewline
trimer nn &  & 1 &  & 1 & 1 &  & 1 & 2 & 2 &  & 1 & \tabularnewline
\hline 
chain loc & 1 &  & 2 &  &  & 2 &  &  &  &  &  & \tabularnewline
chain nn &  & 1 &  & 1 &  &  & 1 & 2 &  &  &  & \tabularnewline
chain nnn &  &  &  &  & 1 &  &  &  & 2 &  &  & \tabularnewline
\hline 
sq. lat.~loc & 1 &  & 4 &  &  & 4 &  &  &  &  &  & \tabularnewline
sq. lat.~nn &  & 1 &  & 1 &  &  & 1 & 6 &  &  &  & 2\tabularnewline
\hline 
tri.~lat.~loc & 1 &  & 6 &  &  & 6 &  &  &  & 6 &  & \tabularnewline
tri.~lat.~nn &  & 1 &  & 1 & 2 &  & 1 & 10 & 4 &  & 2 & 4\tabularnewline
\end{tabular}
\caption{Embedding factor $e(\mathcal{L},i,i^{\prime},g^{(n)}_t)$ for various lattices $\mathcal{L}$ and various separations between external sites $i,i^\prime$. Here 'loc' means local ($i=i^\prime$) while '(n)nn' stands for (next-)nearest neighbor. A missing entry means $e(\mathcal{L},i,i^{\prime},g^{(n)}_t)=0$.
\label{tab:e}
}
\end{table}

Note that only graphs $g^{(n)}$ with a single connected component provide a non-zero contribution to $c_{ii^\prime}^{(n)}(i\nu_m)$. Indeed, if there are multiple connected components, the total 
contribution will vanish in the course of the recursion involved in Eq.~\eqref{eq:c(n)_ii'(m)-rec}. Likewise, we skip graphs with (generalized) leaves, which are parts connected to the spline of the graph (the part with the terminal vertices) by a single edge. Such graphs would only contribute for finite field $h$ which we set to zero in the model \eqref{eq:H}. Similar considerations apply for vacuum graphs to be defined below.

Following these considerations, we have created lists of all required graphs $g^{(n)}$ [with a potential non-zero graph-evaluation $c_{g^{(n)}}(i\nu_m)$], for $n=0,1,...,12$. The number of graphs per order $n \gtrsim 6$ is roughly a factor $4$ larger than for the previous order. At our highest order $n=n_{max}=12$ we need to consider 1273854 graphs $g^{(12)}$. 

Finally we need to consider the quantity which we wanted to evaluate on the lattice snippets defined by a selection of edges $g^{(n)}=\{e_{1},...,e_{n}\}$: The HTE of the Matsubara correlator. We call this HTE the graph evaluation $c_{g^{(n)}}(i\nu_m)$. In Eq.~\eqref{eq:embedding} the correlator expansion on the full lattice is obtained by a sum over all graph evaluations weighted with the embedding factors. In Eq.~\eqref{eq:c_g}, the individual graph evaluations are given by a sum $\sum_{\{e_{1},...,e_{n}\}\rightarrow\{\tilde{e}_{1},...,\tilde{e}_{n}\}}$ which associates the multi-set of the graph's edges to the time-ordered bond operators in $F^{(n)}_{jj^{\prime}}\left(\tilde{e}_{1},...,\tilde{e}_{n};i\nu_{m}\right)$ in all $n!/d[g^{(n)}]$ different ways. Here $d[g^{(n)}]$ yields the graph degeneracy which is the product of factorials of all edge weights.

As in the case of conventional HTE \cite{oitmaaSeriesExpansion2006}, the power of our numerical implementation of Dyn-HTE \cite{Dyn-HTEsoftware_v1_0} rests on the fact that the lattice-specific embedding factors $e(\mathcal{L},i,i^{\prime},g^{(n)})$ can be quickly obtained from advanced graph theoretical algorithms \cite{noauthor_graphsjl_nodate, cordella_subgraph_2004} whereas the numerically costly graph evaluations $ c_{g^{(n)}}(i\nu_m)$ can be pre-computed once and for all. As we will show in the following, they take the form
\begin{equation}
    c_{g^{(n)}_t}(i\nu_m) = \delta_{0,m} c^{(n)}_{t;0} +  \sum_{l=1}^{n} c^{(n)}_{t;l} \; \Delta^{l}_{2\pi m}, 
    \label{eq:c_gl}
\end{equation}
where $\Delta_{2\pi m}\equiv(1-\delta_{0,m})/(2\pi m)$ is the inverse of dimensionless (non-zero) Matsubara frequencies. As we argue later, after the embedding, all odd $l$ vanish on the right-hand side of Eq.~\eqref{eq:embedding} and are thus not computed in the first place. The $c^{(n)}_{t;l}$ for $l=0,2,...,r_{max}$ are rational numbers and together with Eq.~\eqref{eq:TG-expansion} they give rise to the final form \eqref{eq:Dyn-HTE}. In Tab.~\ref{tab:LowOrder-c_gl_1col} we provide the values for the evaluations $c^{(n)}_{t;l}$ for all graphs $g^{(n)}$ with $n \leq 3$ shown in Fig.~\ref{fig:GraphsG}.
The remainder of this section is concerned with the actual calculation of these graph evaluations.  
\begin{table}
    \centering
\begin{tabular}{c|cccccc}
$(n,t)$  & (0,1)  & (1,1)  & (2,1)  & (2,2)  & (2,3)  & (3,1) \tabularnewline
\hline 
$c_{t;0}^{(n)}$  & +1/4  & +1/16  & -1/96  & -1/192  & +1/64  & +1/384 \tabularnewline
$c_{t;2}^{(n)}$  & 0  & 0  & +1/8  & -1/8  & 0  & -1/32 \tabularnewline
\hline 
\hline 
$(n,t)$  & (3,2)  & (3,3)  & (3,4)  & (3,5)  & (3,6)  & (3,7) \tabularnewline
\hline 
$c_{t;0}^{(n)}$  & -1/256  & -1/384  & -1/768  & -1/192  & 0  & +1/256\tabularnewline
$c_{t;2}^{(n)}$  & +1/32  & 0  & 0  & +1/16  & -1/32  & 0\tabularnewline
\end{tabular}
    \caption{Results for the coefficients $c^{(n)}_{t;l}$ in the polynomial \eqref{eq:c_gl} of the evaluation of all graphs $g^{(n)}_t$ in Fig.~\ref{fig:GraphsG} with $n=0,1,2,3$ edges.}
    \label{tab:LowOrder-c_gl_1col}
\end{table}

\subsection{Recursive sub-graph subtractions}

For the (graph-)evaluation of the dynamic spin correlator on graph $g^{(n)}$ denoted by $c_{g^{(n)}}(i\nu_m)$ in Eq.~\eqref{eq:c_g} we are required to assign the multi-set of the graph's edges $\{e_{1},...,e_{n}\}$ to the bond-indices of the operators $\{V_{\tilde{e}_{1}}(\tau_{1}),...,V_{\tilde{e}_{n}}(\tau_{n})\}$ according to $\{e_{1},...,e_{n}\}\rightarrow\{\tilde{e}_{1},...,\tilde{e}_{n}\}$ in all possible ways. The operators are distinguishable by their time arguments even for identical edges. However, recall the iterative definition of the $\mathrm{V-con}$ correlator in Eq.~\eqref{eq:c(n)_ii'(m)-rec} which carries over to the graph evaluation as follows:
\begin{equation}
    c_{g^{(n)}}(i\nu_{m}) = c_{g^{(n)}}^{[full]}(i\nu_{m})-c_{g^{(n)}}^{[sub]}(i\nu_{m}) \label{eq:c[f]-c[sub]}.
\end{equation}
For the first ``full" term, all $n!/d[g^{(n)}]$ assignments are equivalent since the times $\tau_{1},...,\tau_n$ can be relabeled and the order of operators behind $\mathcal{T}$ does not matter. We obtain
\begin{align}
c_{g^{(n)}}^{[full]}(i\nu_{m})=\frac{T^{n+2}}{d[g^{(n)}]}\int_{0}^{\beta}e^{i\nu_{m}(\tau-\tau^{\prime})}\mathrm{d}\tau_{1}...\mathrm{d}\tau_{n}\mathrm{d}\tau\mathrm{d}\tau^{\prime}\nonumber\\
\times\left\langle \mathcal{T}V_{e_{1}}(\tau_{1})...V_{e_{n}}(\tau_{n})S_{j}^{z}(\tau)S_{j^{\prime}}^{z}(\tau^{\prime})\right\rangle_{0}, \label{eq:c[full]}
\end{align}
which we will evaluate further in the next subsection.

\begin{figure}
\begin{centering}
\includegraphics[width=1\columnwidth]{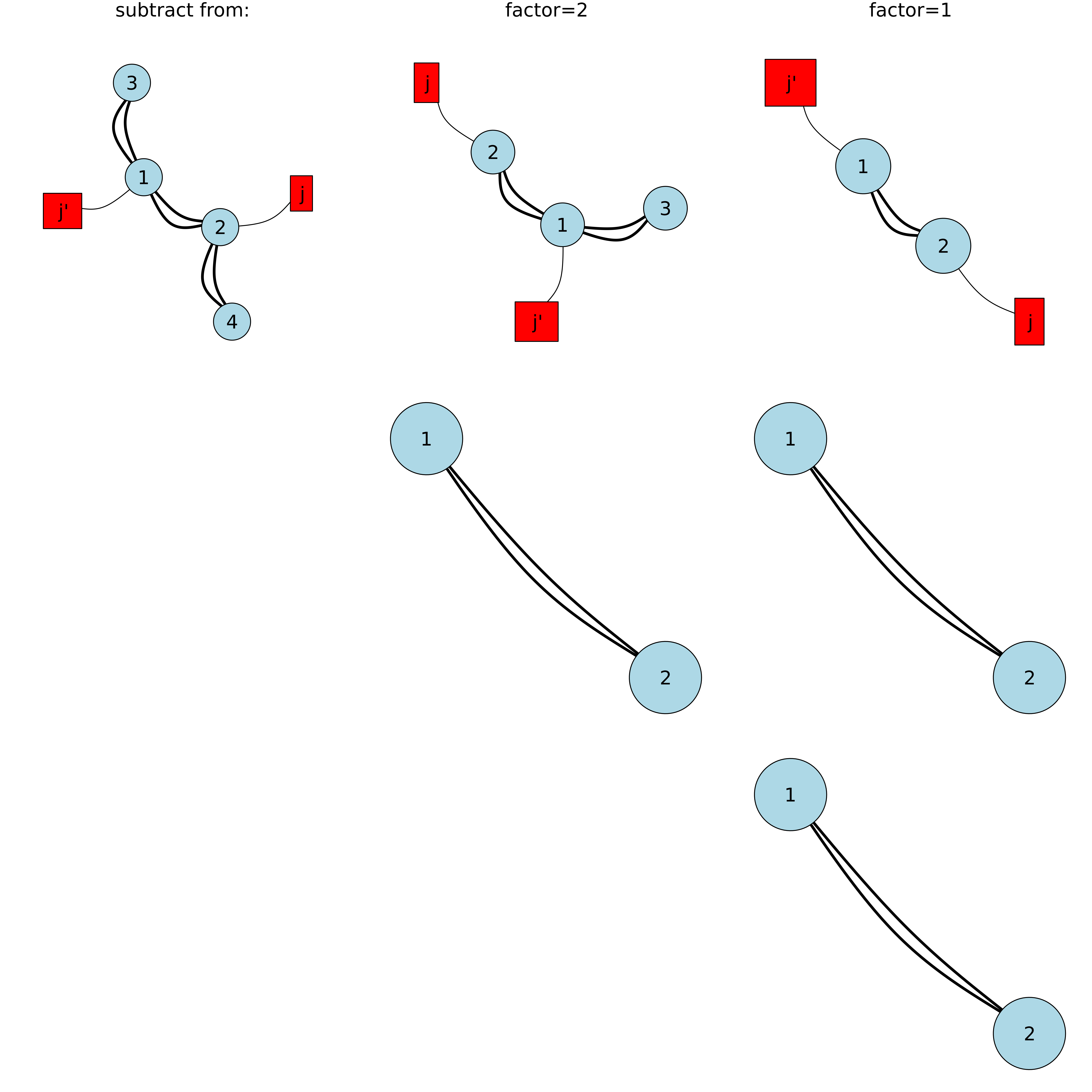}
\par\end{centering}
\caption{\label{fig:VconSubs}
Subtractions from a graph $g^{(6)}$ 
shown in the upper left. All connected sub-graphs $g^{(k)}$ with $k<6$ and non-zero graph evaluation are shown in the top row, second to last column. The vacuum graphs $g^{(6)}\backslash g^{(k)}\equiv v$ are shown in the bottom row. These vacuum graphs are possibly disconnected, see third column. The multiplicity factors $f$ for the particular subtraction are also given.
}
\end{figure}

For the second ``subtracted" term in Eq.~\eqref{eq:c[f]-c[sub]}, we rewrite the $\sum_{S\subsetneq\{1,...,n\}}$ in Eq.~\eqref{eq:c(n)_ii'(m)-rec} as a sum over subgraphs $g^{(k)} \subsetneq g^{(n)}$ with $k<n$ edges. The subgraph $g^{(k)}=\{ e_1^\prime,...,e_k^\prime\}$ still needs to be connected and must contain the terminal vertices $j,j^\prime$. This remaining edges $g^{(n)}\backslash g^{(k)}$ form a vacuum graph without external vertices which is possibly disconnected. There can be more than one ways to get to topologically equivalent $g^{(k)}$, this multiplicity factor is denoted by $f(g^{(n)},g^{(k)})\in\{1,2,...\}$. We refer to Fig.~\ref{fig:VconSubs} for an example listing all possible non-zero subtractions from a particular graph $g^{(6)}$ with $n=6$ edges (left) shown in the 2nd and 3rd column together with their factors $f(g^{(n)},g^{(k)})$ and the $g^{(k)}$ (top), the vacuum graph $g^{(n)}\backslash g^{(k)}$ is shown in the bottom.
Finally, in the second line of Eq.~\eqref{eq:c(n)_ii'(m)-rec} there are $\binom{n}{k}$ possibilities to distribute the imaginary time labels between the first and second average. Hence we obtain
\begin{eqnarray}
    &&c_{g^{(n)}}^{[sub]}(i\nu_{m})	 \label{eq:c[sub]} \\
    &&=
    \frac{\binom{n}{k}}{n!} \!\! \sum_{g^{(k)}\subsetneq g^{(n)}}
    \!\!\!
    f(g^{(n)},g^{(k)})  T^{n+2}
    \!\!
    \int_{0}^{\beta} \!\! e^{i\nu_{m}(\tau-\tau^{\prime})}\mathrm{d}\tau_{1}...\mathrm{d}\tau_{n}\mathrm{d}\tau\mathrm{d}\tau^{\prime} \nonumber \\
	&&\times	 \!\!\!\!
    \sum_{g^{(k)}=\{e_{1}^\prime,...,e_{k}^\prime\}\rightarrow\{\tilde{e}^\prime_{1},...,\tilde{e}^\prime_{k}\}}
    \!\!\!\!\!\!\!\!\!\!\!\!\!\!
    \left\langle \mathcal{T}V_{\tilde{e}^\prime_{1}}(\tau_{1})...V_{\tilde{e}^\prime_{k}}(\tau_{k})S_{j}^{z}(\tau)S_{j^{\prime}}^{z}(\tau^{\prime})\right\rangle _{0}^{\!\mathrm{V-con}} \nonumber \\
	&&\times \!\!\!\!
    \sum_{g^{(n)}\backslash g^{(k)}=\{e^\prime_{k+1},...,e^\prime_{n}\}\rightarrow\{\tilde{e}^\prime_{k+1},...,\tilde{e}^\prime_{n}\}}
    \!\!\!\!\!\!\!\!\!\!\!\!\!\!
    \left\langle \mathcal{T}V_{\tilde{e}^\prime_{k+1}}(\tau_{k+1})...V_{\tilde{e}^\prime_{n}}(\tau_{n})\right\rangle _{0}. \nonumber
\end{eqnarray}
The sum in the penultimate line assigns the multi-set of graph $g^{(k)}$'s  edges $\{e_{1}^\prime,...,e_{k}^\prime\}$ in all possible ways to the bond-indices of the operators
$V_{\tilde{e}^\prime_{1}}(\tau_{1})$,...,$V_{\tilde{e}^\prime_{k}}(\tau_{k})$ and similar in the last line. We now resolve these assignment-sums.

The penultimate line in Eq.\eqref{eq:c[sub]} appears in the graph-resolved  version of Eq.~\eqref{eq:c(m)^(n)} (which only sums over the edges of graph $g^{(k)}$ instead of all bonds $b_1,...,b_n$ of the lattice). Hence we obtain a recursive equation for the graph-evaluation $c_{g^{(n)}}(i\nu_{m})$
\begin{equation}
    c_{g^{(n)}}^{[sub]}(i\nu_{m})
    \! = \!\!
    \sum_{g^{(k)}\subsetneq g^{(n)}}f(g^{(n)},g^{(k)}) \,  c_{g^{(k)}}(i\nu_m) \, \left\langle g^{(n)}\backslash g^{(k)}\right\rangle _{0}. \label{eq:c^sub}
\end{equation}
The last term is the evaluation of the vacuum graph $g^{(n)}\backslash g^{(k)}$ defined from the remaining terms as
\begin{eqnarray}
    &&\left\langle g^{(n)}\backslash g^{(k)}\right\rangle _{0}	 =
    \frac{T^{n-k}}{(n-k)!}\int_{0}^{\beta}\mathrm{d}\tau_{k+1}...\mathrm{d}\tau_{n} \nonumber \\
    &&\times \!\!\!\! \!\!
    \sum_{g^{(n)}\backslash g^{(k)}=
    \{e^\prime_{k+1},...,e^\prime_{n}\}\rightarrow\{\tilde{e}^\prime_{k+1},...,\tilde{e}^\prime_{n}\}}
    \!\!\!\!\!\!
    \left\langle \mathcal{T}V_{\tilde{e}^\prime_{k+1}}(\tau_{k+1})...V_{\tilde{e}^\prime_{n}}(\tau_{n})\right\rangle _{0} \nonumber \\
    &&=
     \frac{1}{d[g^{(n)}\backslash g^{(k)}](n-k)!}\sum_{p\in S_{n-k}}
     \left\langle V_{e^\prime_{k+p_1}}...V_{e^\prime_{k+p_{n-k}}}\right\rangle _{0}. \label{eq:vac-eval}
\end{eqnarray}
In the last step, time-integrals together with the time-ordering $\mathcal{T}$ lead to the sum over all permutations $p$ of the numbers $1,2,...,n-k$, similar to vacuum graph evaluation for conventional HTE. For disconnected vacuum graphs $g^{(n)}\backslash g^{(k)}\equiv v $ which separate into two connected components $v=v_1 \dot{\cup} v_2$ we have $\left\langle v \right\rangle_{0} = \left\langle v_{1}\dot{\cup}v_{2}\right\rangle_{0}=\left\langle v_{1}\right\rangle _{0}\left\langle v_{2}\right\rangle _{0}$ and this generalizes to an arbitrary number of connected components.

In summary, the recursive formulation necessitated by the removal of $V$-disconnected bond operator configurations in the definition \eqref{eq:Vcon} works as follows: Once all the ``full" graph evaluations \eqref{eq:c[full]} at order $n$ and vacuum graph evaluations \eqref{eq:vac-eval} of orders smaller or equal to $n-1$ have been calculated, Eqns.~\eqref{eq:c[f]-c[sub]} and \eqref{eq:c^sub} can be used to find the $n$-th order graph evaluations via the recursion
\begin{eqnarray}
    c_{g^{(n)}}(i\nu_{m}) &=& c_{g^{(n)}}^{[full]}(i\nu_{m}) \label{eq:c_gn-beforeTau} \\
    &-&
    \sum_{g_{k}\subsetneq g_{n}}f(g^{(n)},g^{(k)}) \, c_{g^{(k)}}(i\nu_m) \, \left\langle g^{(n)}\backslash g^{(k)}\right\rangle_{0}.   \nonumber
\end{eqnarray}

This expression is conceptually analogous to a formula popularized in the context of high-order perturbation theory for the fermionic Hubbard model in the diagrammatic Monte-Carlo approach (there an expansion in the Hubbard interaction $U$ is used). In this context, Eq.~\eqref{eq:c_gn-beforeTau} is known as the connected determinant formula and was put forward by Rossi \cite{rossiDeterminantDiagrammatic2017}. In the fermionic problem, the averages $\left\langle \mathcal{T}...\right\rangle _{0}$ corresponding to our ``full" graph evaluations \eqref{eq:c[full]} are taken with respect to a non-interacting fermionic (hopping) Hamiltonian $H_0$ and can thus be obtained via Wick's theorem \cite{bruusManyBodyQuantum2004}. This amounts to a determinant evaluation that is numerically efficient. For canonical bosons, the determinant is replaced by a permanent. Then the time integrals are performed stochastically via Markov-chain Monte-Carlo. 

Crucially, as we are here dealing with spin operators, Wick's theorem is not applicable and the first term on the right-hand side of Eq.~\eqref{eq:c_gn-beforeTau}, $ c_{g^{(n)}}^{[full]}(i\nu_{m})$, given in Eq.~\eqref{eq:c[full]}, cannot be written as a determinant or permanent. On the other hand, it turns out that the time-dependence of the integrand in our case is much simpler than that encountered in the Hubbard model. These observations lead to a dedicated evaluation strategy which we discuss in the next subsection.

\subsection{Kernel trick for imaginary-time integrals}\label{subsec:KernelTrick}

The central quantity left to be computed is $c_{g^{(n)}}^{[full]}(i\nu_{m})$ in Eq.~\eqref{eq:c[full]}, the full part of the graph evaluation. Its $n+2$-fold time integral can be seen as a $n+2$ dimensional temporal Fourier transform of an  (imaginary-)time-ordered $n+2$-point correlator, but with $n$ Matsubara frequencies set to zero. Recently, in Ref.~\onlinecite{halbingerSpectralRepresentation2023} closed-form analytic expressions for these Fourier transforms have been found for arbitrary $n$, generalizing the Lehmann representation \cite{bruusManyBodyQuantum2004} beyond 2-point functions \cite{kuglerMultipointCorrelation2021} (see also \cite{vucicevic_analytical_2021} for an earlier solution of the nested time integrals derived in the context of perturbation theory for the Hubbard model). With the help of these ``kernel functions", the Fourier transform of a time-ordered $n+2$-point correlator is written exactly in terms of many-body eigenstates, eigenenergies and matrix elements of the operators involved in the correlation function \cite{halbingerSpectralRepresentation2023}. 

In a general case the Hamiltonian cannot be diagonalized and the ``kernel trick" is typically not practically applicable. However, in the context of spins, as in Eq.~\eqref{eq:c[full]}, the quantum system of interest is a set of $N$ non-interacting spins ($H_0=0$) with trivial product eigenstates and all eigenenergies vanishing. This allows for a tremendous simplification of the analytical expression for Eq.~\eqref{eq:c[full]} obtained from the kernel trick. For details about the calculation, we refer to our recent work on a complementary spin-diagrammatic scheme tailored for spin systems with close-to-mean-field physics in Ref.~\onlinecite{SchneiderDipolarOrdering2024}.

We obtain our main exact and compact result 
\begin{widetext}
\begin{equation}
    c_{g^{(n)}}^{[full]}(i\nu_{m})=\frac{1}{d[g^{(n)}]}\sum_{p\in S_{n+2}}\tilde{K}_{n+2}(p_{n+1},p_{n+2};m)\left\langle \mathcal{P}V_{e_{1}}^{[p_{1}]}...V_{e_{n}}^{[p_{n}]}S_{j}^{z[p_{n+1}]}S_{j^{\prime}}^{z[p_{n+2}]}\right\rangle _{0} ,
    \label{eq:c_full_final}
\end{equation}
where the index ordering operator $\mathcal{P}$ sorts the operator string according to increasing superscript index (in square-brackets). For example, in the case $n=2$ and for a particular permutation $p$ it acts as $\mathcal{P}V_{e_{1}}^{[4]}V_{e_{2}}^{[1]}S_{j}^{z[3]}S_{j^{\prime}}^{z[2]}=V_{e_{2}}S_{j^\prime}^{z}S_{j}^{z}V_{e_{1}}$.

The dimensionless kernel function $\tilde{K}_{n+2}(p_{n+1},p_{n+2};m)$ carries the dependence on external frequency via the Matsubara integer $m$ and depends on the positions $p_{n+1}$, $p_{n+2}$ of the external operators in the operator string. It relates to the general Kernel function $K_{n+2}$ of Ref.~\onlinecite{halbingerSpectralRepresentation2023} with $n+2$ complex arguments $\Omega_1,...,\Omega_{n+2}$ as
\begin{equation}
    \tilde{K}_{n+2}\left(a_{+},a_{-};m\right) \equiv T^{n+2}K_{n+2}(0,...,0,\underset{\mathrm{pos.}a_{+}}{\underbrace{+i\nu_{m}}},0,...,0,\underset{\mathrm{pos.}a_{-}}{\underbrace{-i\nu_{m}}},0,...,0), \label{eq:def_Ktilde}
\end{equation}
where $a_{\pm}=1,2,...,n+2$ mark the position of the frequency entry $i\nu_{\pm m}$ (of arbitrary relative order $a_{+}\gtrless a_{-})$. In Ref.~\onlinecite{SchneiderDipolarOrdering2024}, we showed that
\begin{equation}
    \tilde{K}_{n+2}\left(a_{+},a_{-};m\right)=\begin{cases}
\frac{1}{(n+2)!} & :m=0,\\
(-1)^{|a_{+}-a_{-}|}\sum_{l=|a_{+}-a_{-}|}^{n+2-\mathrm{min}(a_{+},a_{-})}\left[\Delta_{2\pi mi}\right]^{l}
\frac{\mathrm{\left(sgn[a_{-}-a_{+}]\right)}^{l}}{(n+2-l)!}
\left(\begin{array}{c}
l-1\\
|a_{+}-a_{-}|-1
\end{array}\right) & :m \neq 0,
\end{cases} \label{eq:Ktilde}
\end{equation}
which has the symmetry 
\begin{equation}
\tilde{K}_{n+2}\left(a_{+},a_{-};m\right)=\tilde{K}_{n+2}\left(a_{-},a_{+};-m\right)  
\label{eq:Ktilde_sym}
\end{equation}
required by the definition \eqref{eq:def_Ktilde}. We remark that a somewhat similar treatment for a high-temperature expansion in the context of the fermionic Hubbard model (expansion in $t/T$ with $t$ the hopping-integral) has been put forward in Ref.~\onlinecite{perepelitsky_transport_2016} but to lower orders and also without the geometric flexibility of the graph-based approach. 

In passing, we note a variety of recently suggested approaches that have been developed to solve Matsubara sums (or, equivalently, imaginary-time integrals) that appear in bare or renormalized perturbation theory analytically. These include the discrete-Lehmann representation \cite{Kaye2022DLR}, the intermediate representation \cite{wallerberger2023sparse} and algorithmic Matsubara integration \cite{Taheridehkordi2019AMI,burke2025torchami}. However,  none of these approaches are applicable to perturbation theory diagrams with general bosonic $n$-point correlators as building blocks, like in the present expansion (which can be understood to be of strong-coupling type in Hubbard-model parlance).

\subsection{Computational aspects for ``full" graph evaluations} \label{subsec:computationAspects}

We finally collect a few algorithmic tricks for efficient numerical evaluation of Eq.~\eqref{eq:c_full_final} which turns out to be the performance bottleneck in the calculation of the complete graph evaluation \eqref{eq:c_gn-beforeTau}. This can be skipped by the reader unless interested in details of the numerical implementation or the reason for the absence of odd $l$ in the sum \eqref{eq:c_gl}.
With these tricks in place, the evaluations for all graphs up to and including order $n_{max}=12$ required on the order of $500.000$ core hours for spin length $S=1/2$.

\emph{External operator positions:} The kernel function in Eq.~\eqref{eq:c_full_final} depends only on the positions $p_{n+1}$ and $p_{n+2}$ of the external operators in the operator string, but not on the ordering of the edge operators $V_e$. We thus simplify Eq.~\eqref{eq:c_full_final} by splitting the sum over permutations $p$
into a sum over permutations of the $n$ edge operator positions and
insert the external operator $S_{j}^{z}$ \emph{after} $a$ edge operators
and the other external operator $S_{j^{\prime}}^{z}$ \emph{after} another
$\delta a$ edge operators (and similar for the case that $S_{j^\prime}^z$ appears left of $S_j^z$),
\begin{eqnarray}
c_{g^{(n)}}^{[full]}(i\nu_{m}) & = & \sum_{p\in S_{n}}^{\prime}\sum_{a=0}^{n}\sum_{\delta a=0}^{n-a}\tilde{K}_{n+2}(a+1,a+\delta a+2;m)\left\langle \mathcal{P}V_{e_{1}}^{[p_{1}]}...V_{e_{n}}^{[p_{n}]}\overset{a,\delta a}{\longleftarrow}S_{j}^{z}S_{j^{\prime}}^{z}\right\rangle _{0} \label{eq:c_full_a_Da} \\
 & + & \sum_{p\in S_{n}}^{\prime}\sum_{a=0}^{n}\sum_{\delta a=0}^{n-a}\tilde{K}_{n+2}(a+\delta a+2,a+1;m)\left\langle \mathcal{P}V_{e_{1}}^{[p_{1}]}...V_{e_{n}}^{[p_{n}]}\overset{a,\delta a}{\longleftarrow}S_{j^{\prime}}^{z}S_{j}^{z}\right\rangle _{0}. \nonumber
\end{eqnarray}
The kernels can
be precomputed for any combination of $a,\delta a$ and are only multiplied
with the accumulated traces ($\sim \left\langle...\right\rangle_0$) in the very end. In addition, the primed sum $\sum_{p\in S_{n}}^{\prime}$ excludes equivalent permutations of the multi-set $\{e_{1},...,e_{n}\}$
which in Eq.~\eqref{eq:c_full_final} were canceled by the factor $d[g_{n}]$.

\emph{Even kernel function:} For graphs with $j=j^{\prime}$ the
two traces in the first and second row of Eq.~\eqref{eq:c_full_a_Da} are equivalent,
\begin{equation}
c_{g^{(n)}}^{[full]}(i\nu_{m}) 
=
\sum_{a=0}^{n}\sum_{\delta a=0}^{n-a}
2\tilde{K}_{n+2}^{(even)}(a+1,a+\delta a+2;m)
\sum_{p\in S_{n}}^{\prime}
\left\langle \mathcal{P}V_{e_{1}}^{[p_{1}]}...V_{e_{n}}^{[p_{n}]}\overset{a,\delta a}{\longleftarrow}S_{j}^{z}S_{j^{\prime}}^{z}\right\rangle _{0}\label{eq:c_full_j=jp},
\end{equation}
where the even kernel 
\begin{equation}
    \tilde{K}_{n+2}^{(even)}(a_{+},a_{-};m) \equiv [\tilde{K}_{n+2}(a_{+},a_{-};m)+\tilde{K}_{n+2}(a_{-},a_{+};m)]/2 \label{eq:Keven},
\end{equation}
is given by Eq.~\eqref{eq:Ktilde} with the contributions from odd $l$ removed and is thus real.  

The same simplification is also possible for all other graphs with $j\neq j^{\prime}$.
This builds on the insight that graph evaluations $c_{g^{(n)}}(i\nu_{m})$
are of little practical relevance for themselves and are only required to find $ c_{ii^\prime}^{(n)}(i\nu_m)$ via the embedding in Eq.~\eqref{eq:embedding}. In the latter however, embeddings
with both $(i,i^{\prime})\rightarrow(j,j^{\prime})$ and $(i,i^{\prime})\rightarrow(j^{\prime},j)$
are required. Hence, the rewriting with the even kernel and the vanishing of odd powers of $\Delta_{2\pi m}$ also applies
to this case, but only after the embedding. 
These observations allow us to use Eq.~\eqref{eq:c_full_j=jp} for arbitrary graphs.

\emph{Cyclicity of trace:} Due to the cyclic nature of the trace $\left\langle ...\right\rangle_0$ in Eq.~\eqref{eq:c_full_j=jp}, we can always equate the $p$-sum of the latter for any $a$ by setting $a \rightarrow 0$. Then the sum over $a$ only has to be taken over the kernel,
\begin{equation}
    \tilde{K}_{n+2}^{(even)}(\delta a;m) \equiv \sum_{a=0}^{n-\delta a}\tilde{K}_{n+2}^{(even)}(a+1,a+\delta a+2;m),
\end{equation}
and Eq.~\eqref{eq:c_full_j=jp} simplifies to
\begin{equation}
    c_{g^{(n)}}^{[full]}(i\nu_{m})=2\sum_{\delta a=0}^{n}\tilde{K}_{n+2}^{(even)}(\delta a;m)\sum_{p\in S_{n}}^{\prime}\left\langle S_{j}^{z}\mathcal{P}V_{e_{1}}^{[p_{1}]}...V_{e_{n}}^{[p_{n}]}\overset{\delta a}{\longleftarrow}S_{j^{\prime}}^{z}\right\rangle _{0}.\label{eq:c_full_K_a-summed}
\end{equation}
Anticipating the two embeddings $(i,i^{\prime})\rightarrow(j,j^{\prime})$ and $(i,i^{\prime})\rightarrow(j^{\prime},j)$ as above, we can write the more symmetric expression 
\begin{equation}
    c_{g^{(n)}}^{[full]}(i\nu_{m})=\sum_{\delta a=0}^{n}\left[\tilde{K}_{n+2}^{(even)}(\delta a,m)+\tilde{K}_{n+2}^{(even)}(n-\delta a,m)\right]\sum_{p\in S_{n}}^{\prime}\left\langle S_{j}^{z}\mathcal{P}V_{e_{1}}^{[p_{1}]}...V_{e_{n}}^{[p_{n}]}\overset{\delta a}{\longleftarrow}S_{j^{\prime}}^{z}\right\rangle _{0}. \label{eq:c_full_final_cyclic}
\end{equation}
\end{widetext}
From this equation, also the general form of the expansions in Eq.~\eqref{eq:c_gl} [and \eqref{eq:Dyn-HTE}] follow: From the definition of the kernel \eqref{eq:Ktilde}, after dropping odd powers of $\Delta_{2\pi m}$ in the even kernel, graph evaluations for $g^{(n)}$ with $n$ even contain $\Delta_{2\pi m}^n$ as the highest power. For odd $n$, naively $\Delta_{2\pi m}^{n+1}$ can occur. However, these terms, which emerge only from the $a=0$ contribution to $\tilde{K}_{n+2}^{(even)}(\delta a,m)$ mutually cancel out in the square bracket of Eq.~\eqref{eq:c_full_final_cyclic}.

Regarding algorithmic complexity, note that the sum in Eq.~\eqref{eq:c_full_final_cyclic} has only a factor of $n$ more terms than the analogous expression for the conventional HTE of the equal-time correlator $\left\langle S_j^z S_{j^\prime}^z \right\rangle$ where the two external operators always stay right next to each other due to the absence of time-integrals.

\emph{Flavor sum:} So far we have not considered the structure of the edge operators which according to Sec.~\ref{sec:model} are Heisenberg exchange interactions, $V_{e_{k}}=\sum_{\gamma_{k}}S_{e_{k}(1)}^{\gamma_{k}}S_{e_{k}(2)}^{\bar{\gamma}_k}$. Here the edge-flavor sum is over $\gamma_{k}\in\{+,-,z\}\equiv\{+1,-1,0\}$ and $e_{k}(1)$ and $e_k(2)$ are the two vertices connected by edge $e_k$. As Eq.~\eqref{eq:c_full_final_cyclic} involves $n$ of these edge operators, resolving the $V_{e_{k}}$ for
$k=1,...,n$ naively gives rise to $3^{n}$ flavor combinations to be
summed over. As mentioned below Eq.~\eqref{eq:c(m)^(n)}, the traces $\left\langle ...\right\rangle _{0}$
of strings of equal-time spin operators $\in\{S^{+}_i,S^{-}_i,S^{z}_i\}$ appearing
in Eq.~\eqref{eq:c_full_final_cyclic} factorize into traces of spin
operators with the same vertex (site) index. The required on-site free spin equal-time
correlators for a string of $n^\prime\leq n+2$ spin operators are pre-computed and saved as
a computationally efficient integer after multiplying by a factor $(2S+1)^{n^\prime}$. The product of these factors is to be canceled by global factor $1/(2S+1)^{2n+2}$ which is to be multiplied only at the end. 

It turns
out that the single free spin $n$-point correlators vanish for many operator strings. For example
in the case $S=1/2$ and for spin-operator strings of length $n^\prime = 12$,
only $8172$ out of the $3^{12}=531441$ possible strings have a non-zero
trace. Conditions for a zero trace that do not depend on the operator
order are a breaking of $U(1)$ spin rotation symmetry (unequal numbers of
$S^{+}$ and $S^{-}$ in the string) or a string that is odd under a $\pi$-rotation around the spin-x axis. The latter is the case for a string consisting of an odd number of $S^{z}$ operators, $S^zS^z...S^z$. Such flavor combinations can be skipped
already before entering the $\sum_{p\in S_{n}}^{\prime}$ sum.
We also use the $\pi_{x}$-rotation symmetry in spin space
that ensures equality of the traces for flavor strings connected by
a global exchange of $S^+\leftrightarrow S^-$ operators. 

Once a given permutation $p\in S_{n}$
is selected, operator identities like $(S^{+})^a=0$ for $a>2S$
might render the operator string trivial. Note that this holds also if $S^z$ operators are inserted in between. This is useful since it makes the condition independent of any insertions of external $S^{z}$ operators done in the ${\delta a}$ sum.

\section{Results and applications} \label{sec:Applications}
%
\begin{figure*}
\begin{centering}
\includegraphics{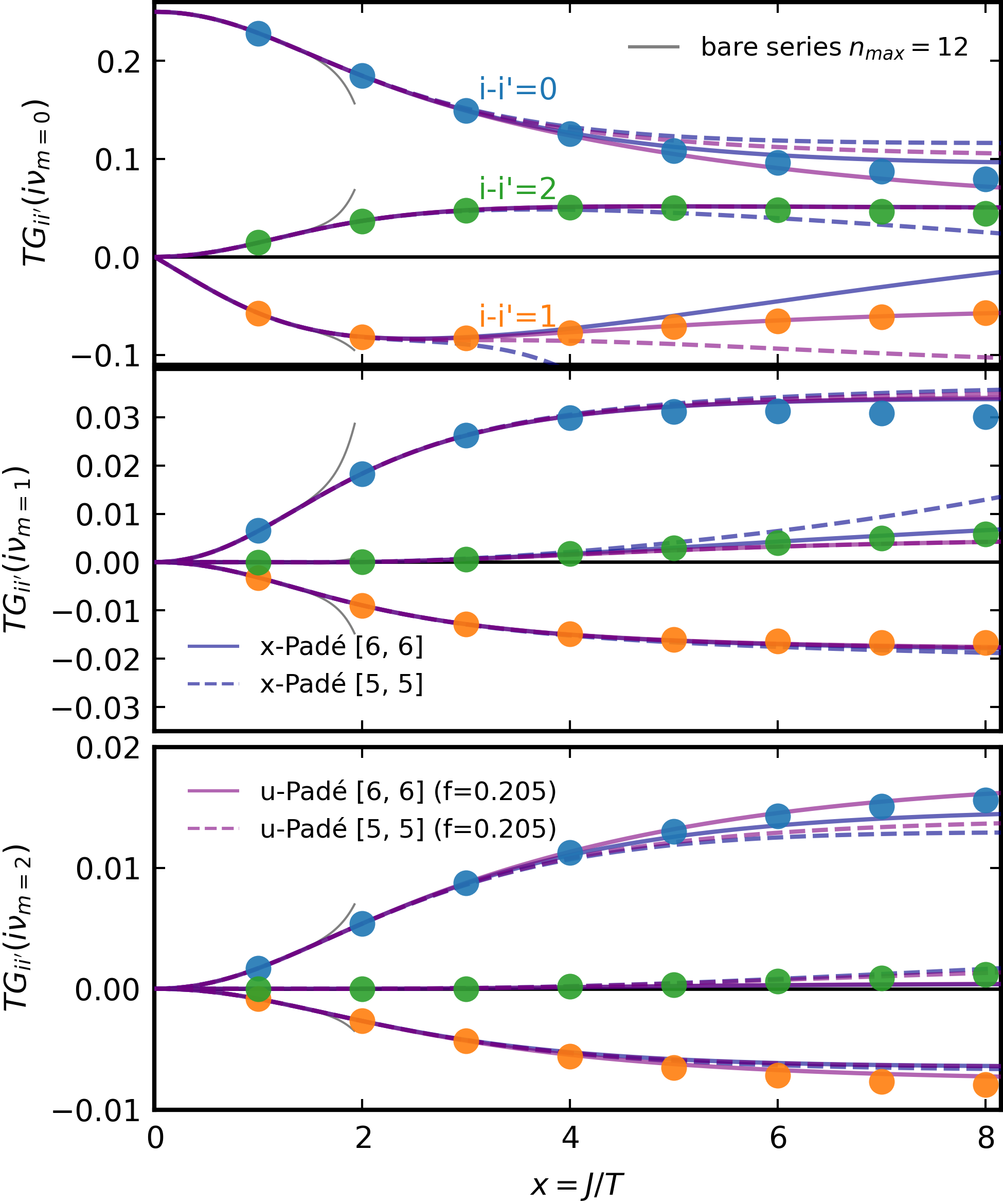}
\hspace{3mm}
\includegraphics{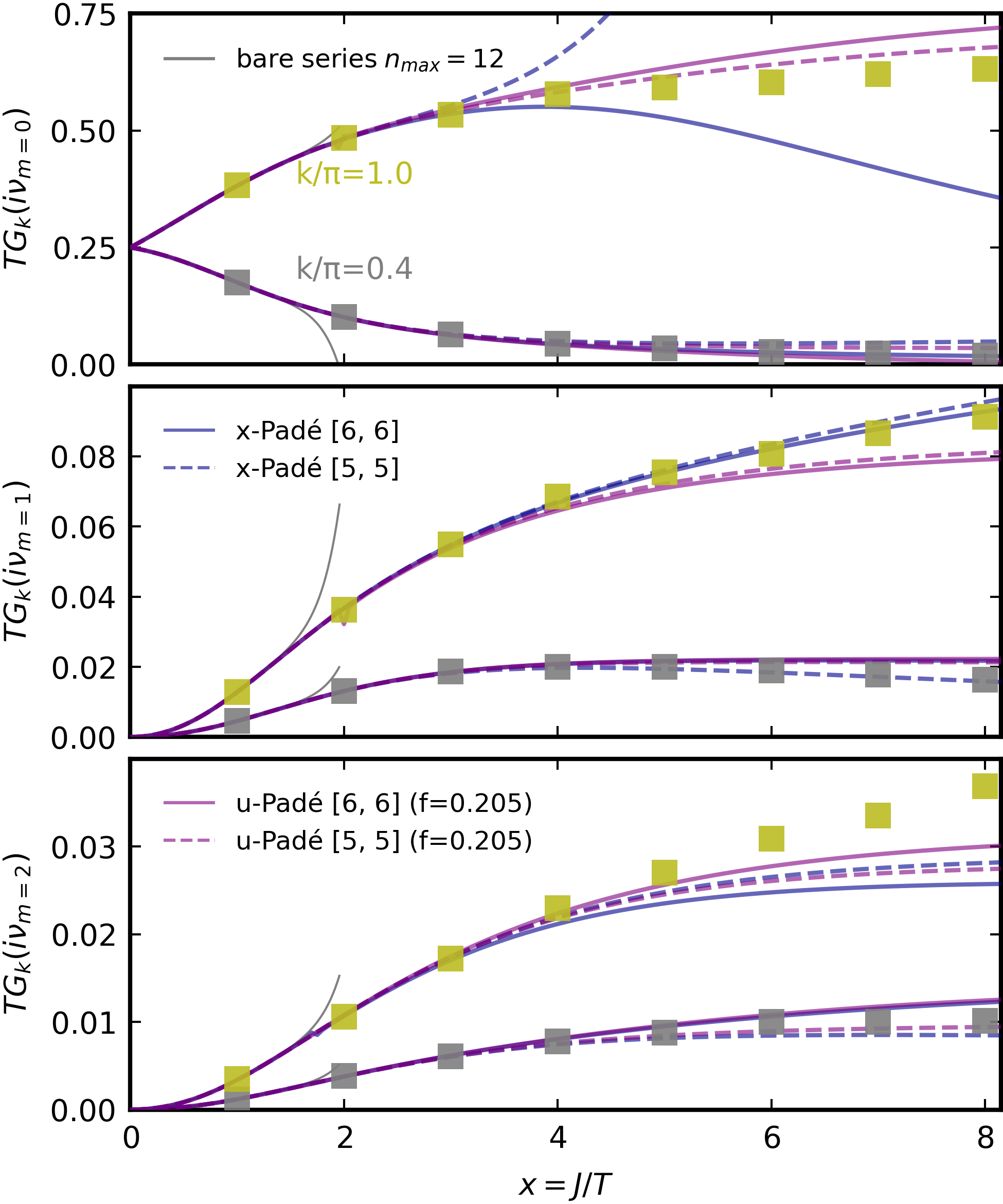}
\par\end{centering}
\centering{}\caption{\label{fig:Chain} 
Matsubara correlators of the Heisenberg $S=1/2$ AFM chain for real-space distances $i-i^\prime=0,1,2$ (left) and wavevectors $k/\pi=0.4,1$ (right) as obtained from Dyn-HTE (lines). The frequencies are $\nu_m=2\pi m T$ with $m=0,1,2$ (top to bottom). Markers show benchmark results from QMC simulations of a 256-site ring with error bars smaller than the symbol size. The thin gray line denotes the evaluation of the bare Dyn-HTE series truncated at order $n_{max}=12$. Symmetric Padé approximants ([6,6] and [5,5]) of the x-series are shown by blue lines (full and dashed), for the transformed series in $u=\mathrm{tanh}(fx)$ with $f=0.205$ they are indicated in purple.
}
\end{figure*}
%

\subsection{Tests of expansion coefficients}

In a first step, we check the rational coefficients in the polynomials $p^{(n)}_{ii^\prime}(x)$ of Eq.~\eqref{eq:Dyn-HTE} obtained from Dyn-HTE in two settings where they are known by other means, see Appendix \ref{app:benchmark_tests} for details. In a first test, we consider the Heisenberg four spin cluster ($N=4$) with all-to-all interactions for which the Matsubara correlator $G_{ii^\prime}(i\nu_m)$ can be found analytically for $S=1/2$, $S=1$ and both for the local and non-local case. The rational coefficients of a series expansion of these exact results in $x=J/T$ agrees to our graph-embedding based Dyn-HTE result. This test thus involves evaluations for \emph{all} graphs with up to four vertices (as they are embeddable in the $N=4$ all-to-all cluster). In a second test we focus on extended lattices in two and three dimensions (like kagome, triangular, pyrochlore, ...) and compute their Dyn-HTE. At each order in $x$, we take the frequency- and site-sums to reproduce the conventional HTE expansion of the (purely static) uniform susceptibility $\chi$ found in the conventional HTE literature. 

\subsection{Benchmark: Heisenberg S=1/2 AFM chain}

We proceed to compute numerical values for the Matsubara correlator \eqref{eq:G(nu_m)} at finite $x=J/T$ based on the Dyn-HTE expansion \eqref{eq:Dyn-HTE}. We start in this subsection  with the (infinite) nearest-neighbor $S=1/2$ Heisenberg AFM chain. We study the Matsubara correlators both in real and momentum space for various distances $i-i^\prime$ and momenta $k$, see Fig.~\ref{fig:Chain}, left and right columns, respectively. Error controlled quantum Monte Carlo (QMC) benchmark data from the worm algorithm \cite{sadouneEfficientScalable2022,lodepolletLodePolletWorm2024} are shown as symbols, error bars are smaller than symbol size. We consider the static case $m=0$ and the two smallest positive Matsubara frequencies $m=1,2$, top to bottom row. 

The summation of the bare Dyn-HTE series up to order $n=12$ (gray line) starts to deviate significantly from the exact results at rather high temperature $x=J/T\simeq2$. This is in close analogy to conventional HTE where it is well known to signal the small radius of convergence of the $x$-series in the complex plane \cite{oitmaaSeriesExpansion2006}. 
A standard tool to extract meaningful information from the bare HTE series at smaller temperatures are Padé approximants \cite{oitmaaSeriesExpansion2006}, which are rational functions ${[K,L]}(x)=\frac{P_{K}(x)}{Q_{L}(x)}$ with polynomials $P_K$ and $Q_L$ of degree $K$ and $L$, respectively. For $K+L = n_{max}$, their coefficients can be determined so that the series expansion of the Padé approximant agrees in the first $n_{max}$ orders with the bare (Dyn-)HTE series. For the case $[K=6,L=6]\equiv [6,6]$, denoted by blue lines in Fig.~\ref{fig:Chain}, the agreement with QMC data already extends to $x \simeq 4$, both for the real- and momentum-space correlator. Note that for the latter, we perform the Fourier transform before computing the Padé approximant. 

For many combinations of Matsubara integer $m$ and distance $i-i^\prime$ (or momentum $k$) the [6,6] Padé approximant agrees with the error controlled QMC result to even much larger $x\simeq 8$, most severe deviations occur for the $m=0$ case with $i-i^\prime = 1$ or $k=\pi$. In real applications (without available benchmark data), the quality of a prediction from Padé approximants is usually gauged by comparison of different $[K,L]$ with $K+L \leq n_{max}$ \cite{oitmaaSeriesExpansion2006}. Due to the asymptotically constant (and finite) values of $TG_{ii^\prime}(i\nu_m)$ at $x\rightarrow \infty$, we limit ourselves to the comparison of the symmetric Padé approximants [6,6] and [5,5] (dashed lines). Indeed, the poor quality of the [6,6] approximants in the two cases mentioned above is reflected in a large disagreement between the two Padé approximants [5,5] and [6,6].

Interestingly, in these cases a transformation from the bare series in $x$ to a series in $u=\mathrm{tanh}(fx)$ and subsequent Padé approximants can help \cite{elstner_spin-12_1994}. Indeed, these u-Padé approximants shown by purple lines in Fig.~\ref{fig:Chain} reasonably agree with the QMC benchmark data down to the smallest temperatures ($x=8$) considered. The free parameter $f=0.205$ is determined from the condition to have the best possible agreement between the [5,5] and [6,6] u-Padés. In principle, $f$ could be chosen differently for each bare series obtained for the set $(m,i-i^\prime)$ [or $(m,k)$], but we picked a global value here for simplicity. Finally, we note that there are also other resummation techniques like variants of Padé, ``integral" approximants \cite{oitmaaSeriesExpansion2006} or conformal maps \cite{bertrand_reconstructing_2019} that could be tested with Dyn-HTE in future work.  

\subsection{Heisenberg S=1/2 AFM on triangular lattice} \label{sec:triangular}
\begin{figure}
\begin{centering}
\includegraphics{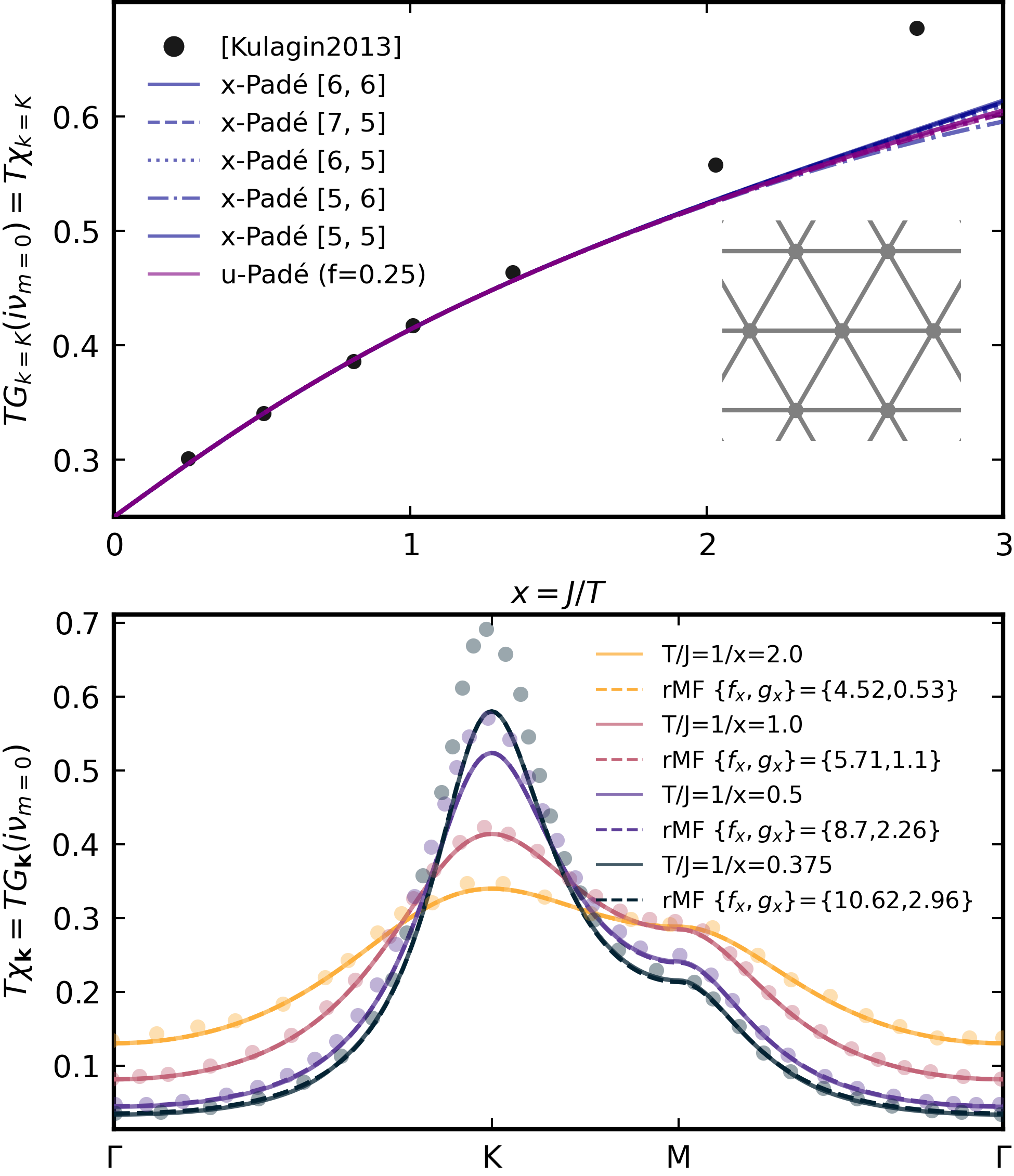}
\par\end{centering}
\centering{}\caption{\label{fig:Triangular} 
Static susceptibility $\chi_\mathbf{k}$ for the Heisenberg $S=1/2$ AFM on the triangular lattice. Results from Dyn-HTE (lines) are compared against the bold-line diagrammatic Monte Carlo (BDMC) data obtained by Kulagin et al.~in Ref.~\cite{kulagin_bold_2013} (dots). Top:  $\chi_{\mathbf{k}=K}$ at the $K$-point in the corner of the hexagonal BZ versus $T$. The convergence of various Padé approximants denoted by blue linestyles is decent and still improves by using the $u$-series (for $f=0.25$, purple lines, same Padé approximants as for $x$-series). Bottom: $\chi_\mathbf{k}$ for a path through the BZ at various $T$. The $\Gamma$-point is the center of the BZ and $M$ denotes the center of the BZ-edge. Away from the $K$-point there is good agreement between BDMC (dots) and Dyn-HTE (full lines). Here the [6,6] $u$-Padé with $f=0.25$ is shown. The dashed lines represent the best fit of the renormalized mean-field form in Eq.~\eqref{eq:QCC} to the Dyn-HTE results. The associated fit parameters $\{f_x,g_x\}$ are given in the legend.
}
\end{figure}
The Heisenberg $S=1/2$ AFM on the triangular lattice serves as an example for a frustrated system for which the Matsubara correlator is difficult to study with QMC due to the sign problem. We focus on the static susceptibility of Eq.~\eqref{eq:chi_isothermal}, $G_\mathbf{k}(i\nu_{m}=0) = \chi_\mathbf{k}$, which has been computed previously using the alternative bold-line diagrammatic Monte Carlo (BDMC) method \cite{kulagin_bold_2013} for spin $S=1/2$. In this approach a fermionic representation is used and standard fermionic skeleton Feynman diagrams are sampled using a Monte Carlo technique. Explicitly, diagrams up to order $7$ in the interaction $J$ are taken into account but the self-consistency condition invokes certain diagram classes at all orders.

In Fig.~\ref{fig:Triangular} we compare the results of Dyn-HTE (lines) and the BDMC (dots). The top panel reports the $T$-dependence of the static susceptibility at the ground-state ordering wavevector at the corner of the hexagonal Brillouin zone (BZ), known as the $K$-point.  The convergence of different Padé approximants denoted by various blue linestyles is decent and still improves by using the $u$-series (with $f=0.25$, purple lines) as in the previous subsection. We are thus confident that the solid purple line (u-Padé [6,6]) gives accurate results in the temperature regime shown with only a few percent error. In contrast, for $x=J/T \geq 2$ the BDMC obtains inconsistent and much larger values. As the BDMC study~\cite{kulagin_bold_2013} does not provide convergence plots of $\chi_{\mathbf{k}=K}$, error bars of the BDMC data are unfortunately not available. As a further complication, the skeleton series is now known for its possibly  convergence to unphysical results \cite{kozik_nonexistence_2015}.
In the bottom panel of Fig.~\ref{fig:Triangular}, we report the static susceptibility for a path through the BZ at various $T$. Away from the $K$-point already detailed in the top panel, there is good agreement between BDMC (dots) and Dyn-HTE (solid lines) for all temperatures studied.

\subsection{Renormalized mean-field form of static susceptibility}

In recent work \cite{schneiderTamingSpin2024} two of us showed that the static susceptibility $\chi_\mathbf{k} = G_\mathbf{k}(i\nu_{m}=0)$ of a wide selection of Heisenberg models on lattices in dimension higher than one can very well be approximated by a two-parameter function of $\mathbf{k}$, the \textit{renormalized mean-field} form (rMF). The standard mean-field expression for nearest-neighbor models with coupling $J$ reads  
\begin{align}
    [TG_{\mathbf{k}}^{\mathrm{MF}}(i\nu_{m}=0)]^{-1} = 1/b_1+ x \gamma_1(\mathbf{k}), \label{eq:inv_MF}
\end{align}
where $b_1 = S(S+1)/3$ denotes the dimensionless susceptibility of a free spin and $\gamma_n(\mathbf{k})= \sum_{\mathbf{r} \in n\mathrm{th \, NN} }\exp(i\mathbf{k}\cdot\mathbf{r})$ is the spatial Fourier transform of the $j$-th nearest-neighbor coupling pattern [Eq.~\eqref{eq:inv_MF} only involves $\gamma_1$].  
In the non-symmetry broken regime, the $\mathbf{k}$-dependence of the exact static susceptibility can always be expanded by using the remaining $\gamma_n(\mathbf{k})$ with $n>1$,
\begin{align}
     [TG_{\mathbf{k}}(i\nu_{m}=0)]^{-1} &= f_x+g_x\gamma_1(\mathbf{k})+\epsilon_2 \gamma_2(\mathbf{k}) + \ldots \label{eq:susc_exact} \\
     &\approx f_x+g_x\gamma_1(\mathbf{k}). \label{eq:QCC} 
\end{align}
The approximation in Eq.~\eqref{eq:QCC} is the rMF form with parameters $f_x$ and $g_x$ replacing $1/b_1$ and $x$ from Eq.~\eqref{eq:inv_MF}, respectively. The expansion in Eq.~\eqref{eq:susc_exact} is defined by inverse Fourier transforms with respect to real-space vectors $\mathbf{r}_{n \mathrm{th \, NN}}$ pointing to the $n$th nearest-neighbor,
\begin{align}
    f_x &= \frac{1}{N_b V_{\mathrm{BZ}}}\int_{\mathrm{BZ}}\mathrm{d}\mathbf{k}[TG_{\mathbf{k}}(i\nu_{m}=0)]^{-1},\label{eq:QCC_f}\\
    g_x &= \frac{1}{N_b V_{\mathrm{BZ}}}\int_{\mathrm{BZ}}\mathrm{d}\mathbf{k} e^{i \mathbf{k}\cdot \mathbf{r}_{\mathrm{NN}}}[TG_{\mathbf{k}}(i\nu_{m}=0)]^{-1}\label{eq:QCC_g},\\
    \epsilon_{n} &= \frac{1}{N_b V_{\mathrm{BZ}}}\int_{\mathrm{BZ}}\mathrm{d}\mathbf{k} e^{i \mathbf{k}\cdot \mathbf{r}_{n \mathrm{th \, NN}}}[TG_{\mathbf{k}}(i\nu_{m}=0)]^{-1}. \label{eq:QCC_eps}
\end{align}
Here $V_{\mathrm{BZ}}$ is the BZ volume and $N_b$ the number of basis sites.

In Ref.~\onlinecite{schneiderTamingSpin2024} we could show analytically that the HTE of the beyond-rMF terms $\epsilon_n$ for $n=2,3,...$ start at order $O(x^4)$ with a particularly small prefactor. This lead us to conjecture the excellent validity of the rMF beyond the high-temperature regime which was corroborated by matching the rMF form \eqref{eq:QCC} with published momentum-dependent susceptibilities obtained from the BDMC method \cite{kulagin_bold_2013} and pseudo-fermion functional renormalization group \cite{gonzalez_dynamics_2024}. 

Dyn-HTE now offers an alternative and more reliable source of static susceptibility data.
For $\chi_\mathbf{k}$ of the triangular lattice model shown in Fig.~\ref{fig:Triangular}(bottom) we determined the rMF parameters $f_x$ and $g_x$ of Eq.~\eqref{eq:QCC} via least-squares fit to the Dyn-HTE data (see legend and dashed lines). The rMF form of the susceptibility describes the Dyn-HTE data remarkably well, with an almost perfect match even for the lowest temperature $T/J = 0.375$.
\begin{figure}
    \centering
    \includegraphics[width=\linewidth]{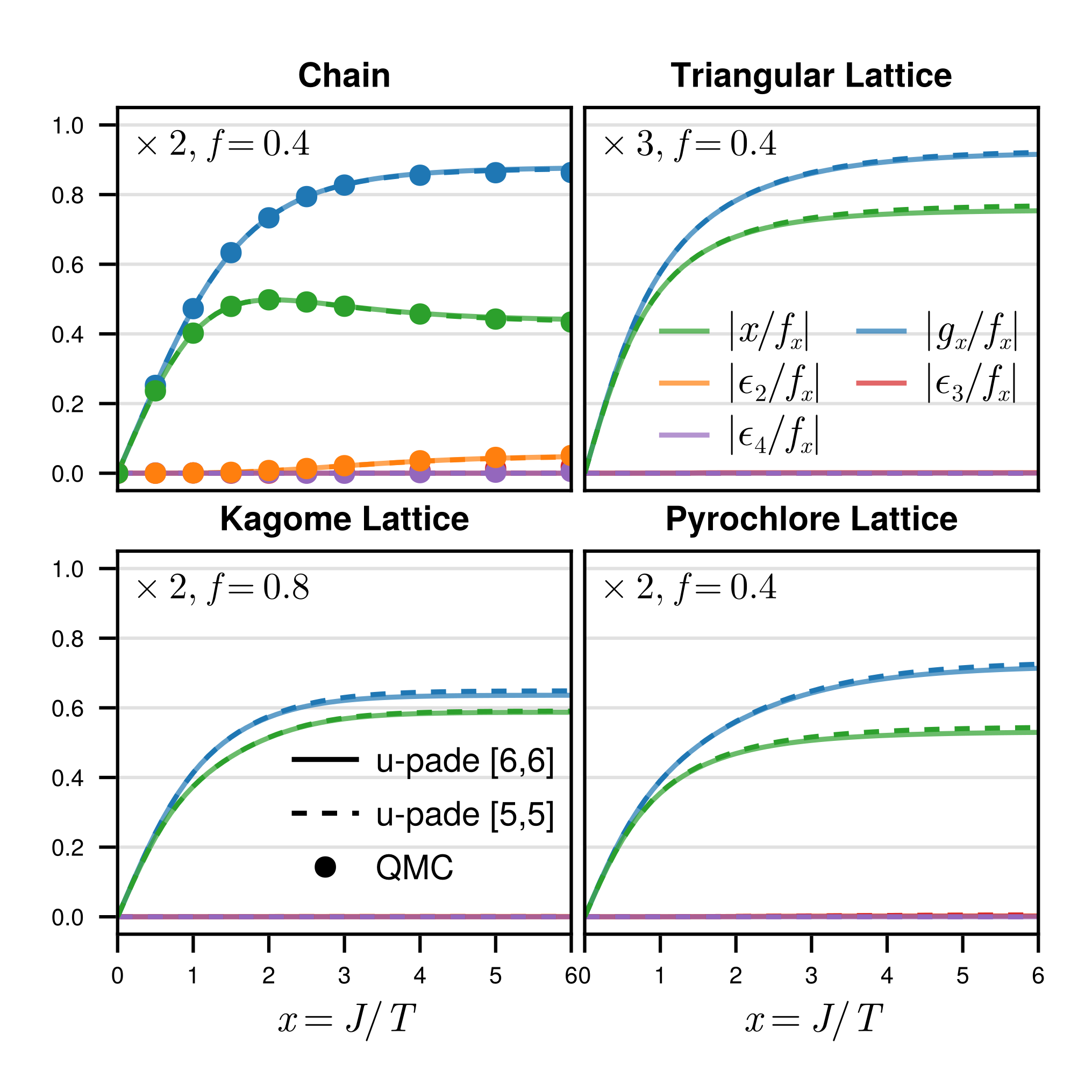}
    \caption{Amplitudes of parameter-ratios of the rMF form \eqref{eq:QCC}: $x/f_x$, $g_x/f_x$ and the first few corrections $\epsilon_{2,3,4}/f_x$ [c.f.~Eq.~\eqref{eq:susc_exact}] for the nearest-neighbor $S=1/2$ Heisenberg AFM model on the chain, triangular, kagome and pyrochlore lattice from Dyn-HTE. For resummation, we used $[6,6]$ and $[5,5]$ u-Padé approximants. The data is multiplied by the maximum (eigenvalue) of $-\gamma_1(\mathbf{k})$ given in the panels alongside the u-Padé parameter $f$. For the chain geometry QMC data (dots) \cite{lodepolletLodePolletWorm2024} matches well with the Dyn-HTE data (lines). For triangular, kagome and pyrochlore lattice $\epsilon_{2,3,4}/f_x$ are of order $10^{-3}$ rendering the rMF an excellent approximation. Whenever the [6,6] or [5,5] Padé approximant is anomalous (with obvious poles), we use a stable lower-order Padé approximant.}
    \label{fig:rMF}
\end{figure}

A quantitative measure of the rMF form's accuracy is provided by the ratio $|\frac{\sum_n\epsilon_n\gamma_n(\mathbf{k})}{f_x + g_x \gamma_1(\mathbf{k})}|$ for which values much smaller than unity indicate the applicability of the approximation. 
In Fig.~\ref{fig:rMF} we report several ratios of $f_x$, $g_x$ and the numerically dominant $\epsilon_2,\epsilon_3,\epsilon_4$ for the Heisenberg $S=1/2$ AFM on various lattices. They are obtained from the $u$-Padé approximations for the $x$-series expansions of the respective ratio found from Dyn-HTE. For the chain (top left panel), the ratios $1/f_x$, $g_x/f_x$ and $\epsilon_n/f_x$ saturate for low $T$ (large $x$), hence resummation of the ratios with the Padé approximant of the $u$-series works very well as witnessed by comparison to QMC (dots). As $\epsilon_2$ grows sizably for low $T$, the validity of the rMF approximation is compromised. This was stated before in Ref.~\onlinecite{kulaginBoldDiagrammatic2013}.

However, for the two and three dimensional systems (triangular, kagome and pyrochlore lattice), the results in Fig.~\ref{fig:rMF} indicate that $\epsilon_2,\epsilon_3,\epsilon_4$ are negligible as compared to $f_x$ and $g_x$ underlining the validity of the rMF approximation. Remarkably, $\epsilon_n$ stays at least two
orders of magnitude lower than $g_x$ over the whole accessible temperature range. Therefore, the static susceptibility is accurately described by just $g_x$ and $f_x$ for these models. This also explains the excellent fit to the rMF form in Fig.~\ref{fig:Triangular}. Other corrections $\epsilon_{n}$ for $n>4$ (not shown) stay very small as well.

\section{Conclusion and outlook}

We have presented a dynamical extension of the high-temperature series expansion (HTE) to the Matsubara spin correlator for spin Hamiltonians (Dyn-HTE). Currently, our numerical implementation \cite{Dyn-HTEsoftware_v1_0} features an expansion to order $n_{max}=12$ and is available for $S \leq 1$ Heisenberg Hamiltonians with a single coupling constant $J$ on arbitrary (in particular frustrated and high-dimensional) lattices. The real-frequency dynamic structure factor can be obtained via postprocessing of the Matsubara data as described in our companion letter \cite{burkard_DynHTE_letter}.

For further methodological development of Dyn-HTE it would be worthwhile to reduce the number of graphs by the free-graph expansion technique (where different vertices of a graph can be assigned to the same or different lattice sites) or by the linked-cluster method \cite{oitmaaSeriesExpansion2006}. In the latter, only graphs (clusters) with simple edges are required on which the observable needs to be expanded in $x=J/T$ or is even determined by exact diagonalization. While this could be a viable way to higher expansion orders for small $S$, it would also be interesting to extend Dyn-HTE in the form presented in this work to arbitrary $S$. This could be achieved by building on closed-form expressions for the equal-time free spin correlators \cite{halbingerSpectralRepresentation2023}.  

Moreover, possible extensions of Dyn-HTE analogous to achievements in conventional HTE include the application to Heisenberg models with more than one coupling constant, e.g.~$J_1$-$J_2$ models \cite{hehnHightemperatureSeries2017}, single-ion anisotropies or magnetic fields \cite{pierre_high_2024}. In the latter case, the kernel trick for the evaluation of the imaginary-time integrals needs to be adapted for the presence of (a few) non-zero many-body eigenenergies of $H_0$. It would also be interesting to aim Dyn-HTE at more complex correlation functions like three-point correlators encoding higher-order response, see Ref.~\cite{ruckriegel_recursive_2024} for pioneering work. Likewise, it would be useful to improve the series convergence beyond the standard Padé approximants by implementing advanced ideas like, e.g., the homotopic action \cite{kim_homotopic_2021}.

Beyond the Heisenberg case, Dyn-HTE can be extended to models with broken or reduced spin rotation symmetry like the XXZ case. Likewise, also the fermionic (or bosonic) $t$-$J$-model which is prominently realized in cold-atom quantum simulation \cite{koepsell_microscopic_2021} should be considered. Here, existing treatments \cite{metznerLinkedclusterExpansionAtomic1991,perepelitsky_transport_2016} could be extended to higher expansion orders.

\section{Acknowledgment}
We acknowledge useful discussions with Michel Ferrero at several stages of the project and are grateful to Lode Pollet for advise on the QMC worm algorithm \cite{lodepolletLodePolletWorm2024}. We further thank Nikolay Prokofiev and Boris Svistunov for useful remarks on the manuscript.

The authors acknowledge the Gauss Centre for Supercomputing e.V. (www.gauss-centre.eu) for funding this project by providing computing time through the John von Neumann Institute for Computing (NIC) on the GCS Supercomputer JUWELS at Jülich Supercomputing Centre (JSC). The authors also acknowledge support by the state of Baden-Württemberg through bwHPC and the German Research Foundation (DFG) through Grant No.~INST 40/575-1 FUGG (JUSTUS 2 cluster).
We acknowledge funding from the Deutsche
Forschungsgemeinschaft (DFG, German Research Foundation) through the Research Unit FOR 5413/1, Grant No.~465199066. B.Sch.~acknowledges funding from the Munich
Quantum Valley, supported by the Bavarian state government
with funds from the Hightech Agenda Bayern Plus. B.Sb. and
B.Sch.~are supported by DFG Grant No.~524270816.

\appendix
\onecolumngrid

\section{Benchmark tests for expansion of Matsubara correlator} 
\label{app:benchmark_tests}

\subsection{Small spin cluster: N=4 spins with all-to-all coupling}

As a first benchmark check on the expansion coefficients of Dyn-HTE in Eq.~\eqref{eq:Dyn-HTE}, we consider a cluster with $N=4$ spins coupled by all-to-all interaction $J$, for both $S=1/2$ and $S=1$. We compute the exact local and non-local Matsubara correlators \eqref{eq:G(nu_m)} in closed form using diagonalization and the spectral (Lehmann) representation \cite{bruusManyBodyQuantum2004}. For $S=1/2$ they read
\begin{eqnarray}
TG_{11}({i\nu_{m}})	
&\!=\!&	\begin{cases} \!
\frac{5 (x-1)+e^{2 x} \left(9 x+8 e^x-3\right)}{8 \left(e^{2 x} \left(2 e^x+9\right)+5\right) x}
=
\frac{1}{4}-\frac{x^2}{32}+\frac{x^3}{128}+\frac{23 x^4}{2560}+... & :m\!=\!0,\\

\frac{\Delta ^2 \left(e^x-1\right) x \left(5 \Delta ^2 x^2+e^x \left(5 \Delta ^2 x^2+e^x \left(8 \Delta ^2 x^2+2\right)+5\right)+5\right)}{2 \left(e^{2 x} \left(2 e^x+9\right)+5\right) \left(4 \Delta ^4 x^4+5 \Delta ^2 x^2+1\right)}
= 
\! \frac{3 \Delta ^2 x^2}{8} \!-\! \frac{3 \Delta ^2 x^3}{32}
\!-\! \left(\frac{21 \Delta ^4}{16} \!+\! \frac{11 \Delta ^2}{128}\right) x^4
\!+... & :m\!\neq\!0,
\end{cases} \\
TG_{12}({i\nu_{m}})	&\!=\!&	\begin{cases} \!
\frac{15 x+e^{2 x} \left(3 x-8 e^x+3\right)+5}{24 \left(e^{2 x} \left(2 e^x+9\right)+5\right) x} 
=
-\frac{x}{16}+\frac{5 x^2}{192}+\frac{3 x^3}{256}-\frac{221 x^4}{15360}+...
& \!\!:\!m\!=\!0,\\
\frac{\Delta ^2 \left(e^x-1\right) x \left(5 \Delta ^2 x^2+e^x \left(5 \Delta ^2 x^2+e^x \left(8 \Delta ^2 x^2+2\right)+5\right)+5\right)}{-6 \left(e^{2 x} \left(2 e^x+9\right)+5\right) \left(4 \Delta ^4 x^4+5 \Delta ^2 x^2+1\right)}
= -\frac{\Delta ^2 x^2}{8} +\frac{\Delta ^2 x^3}{32}\!+\!\left(\frac{7 \Delta ^4}{16}+\frac{11 \Delta ^2}{384}\right) x^4+...\!\!
& \!\!:m\!\neq\!0.
\end{cases}
\end{eqnarray}
Here we abbreviate $\Delta=\Delta_{2\pi m}$. The straightforward series expansions in $x$ (shown here to order $x^4$ for brevity) are reproduced via Dyn-HTE based on Eqns.~\eqref{eq:TG-expansion}, \eqref{eq:embedding}, \eqref{eq:c_gn-beforeTau} and \eqref{eq:c_full_final} up to the maximum order $n_{max}=12$. The same holds for the case $S=1$ (the full expressions for $m\neq 0$ are too long to be shown)
\begin{eqnarray}
TG_{11}({i\nu_{m}})	
&\!=\!&	\begin{cases} \frac{270 x+288 e^{10 x}+108 e^{9 x} (3 x+1)+20 e^{7 x} (33 x-7)+7 e^{4 x} (66 x-25)-81}{216 \left(7 e^{4 x}+10 e^{7 x}+6 e^{9 x}+e^{10 x}+3\right) x}
= \frac{2}{3} -\frac{2 x^2}{9} + \frac{13 x^3}{54} + \frac{181 x^4}{1620}
+... & :m\!=\!0,\\
\frac{8 \Delta ^2 x^2}{3}
-\frac{26 \Delta ^2 x^3}{9}
+\left(-\frac{244 \Delta ^4}{9}-\frac{8 \Delta ^2}{9}\right) x^4
+... & :m\!\neq\!0,
\end{cases} \\
TG_{12}({i\nu_{m}})	&\!=\!&	\begin{cases} \! 
\frac{810 x-288 e^{10 x}-108 e^{9 x} (x+1)+140 e^{7 x} (3 x+1)+175 e^{4 x} (6 x+1)+81}{648 \left(7 e^{4 x}+10 e^{7 x}+6 e^{9 x}+e^{10 x}+3\right) x}
= 
-\frac{4 x}{9}
\!+\frac{5 x^2}{9}
\!+\frac{11 x^3}{162}
\!-\!\frac{3503 x^4}{2430}
+...
& \!\!:\!m\!=\!0,\\
-\frac{8 \Delta ^2 x^2}{9}
+\frac{26 \Delta ^2 x^3}{27}
+\left(\frac{244 \Delta ^4}{27}+\frac{8 \Delta ^2}{27}\right) x^4
+...\!\!
& \!\!:m\!\neq\!0.
\end{cases}
\end{eqnarray}

\subsection{Consistency with conventional HTE for uniform susceptibility}

As mentioned above, the conventional HTE for equal-time correlators and thermodynamic quantities like the uniform susceptibility $\chi  \equiv \sum_{i^\prime} \left\langle S_{i}^{z}S_{i^{\prime}}^{z}\right\rangle$ is well established. Here we test Dyn-HTE by reproducing the expansion coefficients of $\chi$ calculated to high orders for many lattices. First, we recover the equal-time correlators from the Matsubara correlators by a frequency sum,
\begin{equation}
    \left\langle S_{i}^{z}S_{i^{\prime}}^{z}\right\rangle =\sum_{m}TG_{ii^{\prime}}\left(i\nu_{m}\right). \label{eq:G_ii'_m_sum}
\end{equation} 
We insert the Dyn-HTE expansion from Eq.~\eqref{eq:TG-expansion} on the right-hand side and perform the frequency sum. The numerical values for the sum over even powers of $\Delta_{2\pi m}$ are summarized in Tab.~\ref{tab:Frequency_Sums} for convenience.

As a first non-trivial consistency check, we calculate the uniform susceptibility $\chi$ for the triangular lattice Heisenberg $S=1/2$ AFM and reproduce the expansion coefficients provided in Ref.~\onlinecite{OitmaaHTESpinHalfHeisenberg1996} for all orders $n \leq n_{max}=12$. Note that this uses all graph evaluations for graphs embeddable in the triangular lattice with their \emph{complete} frequency dependence [only a static contribution must survive after the sum in Eq.~\eqref{eq:G_ii'_m_sum}] and thus constitutes a rather non-trivial check.

For the Kagome lattice, we also reproduce the HTE coefficients for the susceptibility~\cite{elstner_spin-12_1994}. The $n$-th order coefficient of $\chi$ in $(-x)$ is given by $a_{n+1}(-1)^n /[(n+1)! \: 4^{n+2}]$ where $a_n$ is given in the right column of Tab.~I in Ref.~\onlinecite{elstner_spin-12_1994}. There, a misprint must be corrected, $a_7=2 711 296$. This has been also noticed in Ref.~\onlinecite{lohmannTenthorderHightemperature2014}. For $S=1$ we performed the same check for $\chi$ against the conventional HTE~\cite{lohmannTenthorderHightemperature2014}. 

Finally, for the Heisenberg $S=1/2$ AFM on the pyrochlore lattice in three dimensions, we reproduced the HTE coefficients published in Ref.~\cite{gonzalez_finite-temperature_2023}.
\begin{table}[h!]
\centering
{\renewcommand{\arraystretch}{1.3}
\begin{tabular}{c|c c c c c c c c c }
$l$ & $2$ & $4$ & $6$ & $8$ & $10$ & $12$ & $14$ & $16$ & $18$\tabularnewline
\hline  
$\Sigma_{m}\Delta_{2\pi m}^{l}$ & $\frac{1}{12}$ & $\frac{1}{720}$ & $\frac{1}{30240}$ & $\frac{1}{1209600}$ & $\frac{1}{47900160}$ & $\frac{691}{1307674368000}$ & $\frac{1}{74724249600}$ & $\frac{3617}{10670622842880000}$ & $\frac{43867}{5109094217170944000}$\tabularnewline
\end{tabular} 
}
\caption{Frequency sums needed in Eq.~\eqref{eq:G_ii'_m_sum}.} \label{tab:Frequency_Sums}
\end{table}

\twocolumngrid
\bibliography{DynHTE.bib}

\end{document}